\documentclass[aps,prl,reprint,superscriptaddress]{revtex4-1}
\usepackage{amsmath}
\usepackage{graphicx}
\usepackage{amsfonts}
\usepackage{color}
\usepackage{upgreek}

\newcommand{\Ham}{\hat{\mathcal{H}}}

\makeatletter
\def\captionof#1#2{{\def\@captype{#1}#2}}
\makeatother
\usepackage[T1]{fontenc}
\usepackage{graphicx,epsfig}
\usepackage{epstopdf}
\usepackage{mathrsfs}
\usepackage{amsmath,amssymb,bbm,bm}
\usepackage{float}
\usepackage{import}
\usepackage{booktabs}
\usepackage{exscale}
\usepackage[english]{babel}
\usepackage{color}
\usepackage{appendix}
\usepackage{dsfont}
\usepackage{hyperref}
\hypersetup{
    colorlinks,
    linkcolor={rgb:red,2; blue,1},
    citecolor={blue!60!black},
    urlcolor={blue!60!black}
}
\usepackage{xcolor}

\newcommand{\ii}{\ensuremath{\mathrm{i}}}
\newcommand{\ee}{\ensuremath{\mathrm{e}}}
\newcommand{\dd}{\mathrm{d}}

\newcommand{\dw}[1]{\ensuremath{\frac{\mathrm{d}#1}{2\pi}}}

\renewcommand{\Re}{\,\mathrm{Re}\,}

\newcommand{\Tr}{\mathrm{Tr}}

\renewcommand{\b}{\ensuremath{\hat{b}}}
\newcommand{\bdag}{\ensuremath{\hat{b}^\dagger}}
\renewcommand{\c}{\ensuremath{\hat{c}}}

\newcommand{\sg}{\ensuremath{\hat{\sigma}}}

\begin{document}

\title{Tunable phonon induced steady state coherence in a double quantum dot}

\author{Archak Purkayastha}
\email{archak.p@tcd.ie}
\affiliation{School of Physics, Trinity College Dublin, College Green, Dublin 2, Ireland}
\author{Giacomo Guarnieri}
\email{guarnieg@tcd.ie}
\affiliation{School of Physics, Trinity College Dublin, College Green, Dublin 2, Ireland}
\author{Mark T. Mitchison}
\email{mark.mitchison@tcd.ie}
\affiliation{School of Physics, Trinity College Dublin, College Green, Dublin 2, Ireland}
\author{Radim Filip}
\affiliation{Department of Optics, Palack\'{y} University, 17. listopadu 1192/12, 771 46 Olomouc, Czech Republic}
\author{John Goold}
\email{gooldj@tcd.ie}
\affiliation{School of Physics, Trinity College Dublin, College Green, Dublin 2, Ireland}
  
\date{\today}

\begin{abstract}
Charge qubits can be created and manipulated in solid-state double-quantum-dot (DQD) platforms. Typically, these systems are strongly affected by quantum noise stemming from coupling to substrate phonons. This is usually assumed to lead to decoherence towards steady states that are diagonal in the energy eigenbasis. In this article we show, to the contrary, that due to the presence of phonons the equilibrium steady state of the DQD charge qubit spontaneously exhibits coherence in the energy eigenbasis with high purity. The magnitude and phase of the coherence can be controlled by tuning the Hamiltonian parameters of the qubit. The coherence is also robust to presence of fermionic leads. In addition, we show that this steady-state coherence can be used to drive an auxiliary cavity mode coupled to the DQD.
\end{abstract}

\maketitle

Preparation and coherent control of qubit states is at the heart of many quantum technologies~\cite{nielsen2002quantum}. Undoubtedly, quantum coherence is the most primordial non-classical effect and is the root of many advantages displayed by quantum technologies over their classical equivalents. One of the major challenges for applications is to prepare a qubit in a state which has a controllable amount of coherence with stability in the long-time limit~\cite{bennett2000quantum}. In real physical systems, quantum coherence is usually a fragile property, which is eventually destroyed by the presence of a surrounding environment~\cite{Alicki2007, Breuer2002}. It is therefore no surprise that a plethora of strategies to preserve coherence have been conceived, such as quantum error correction~\cite{Shor1995}, dynamical decoupling~\cite{Viola1999} or feedback control~\cite{Rabitz2005}. All of these schemes are to some degree an inevitable battle against decoherence. Rather than fighting this battle, here we highlight a different counter-intuitive route to generate and preserve coherence using quantum noise.

We specifically consider a semiconductor double-quantum-dot (DQD) embedded on a substrate~\cite{measuring_phonons_2010,Colless2014,Petta1,Petta2,Petta3,Petta4,Petta5} which realize a charge qubit coupled to a phononic bath ~\cite{DiVincenzo,phonon_spectral_deriv}. It has been previously demonstrated that properties of the DQD may be used to extract information about the phononic bath \cite{measuring_phonons_2010,Colless2014}. In a recent experiment, ~\cite{Petta1, Petta2}, the DQD and substrate were coupled to an auxiliary optical cavity which was in turn used to experimentally characterize the spectral density of the substrate phonons. In this article, we model the dynamical evolution of these platforms and focus on their steady state-properties. Remarkably, we find that the presence of phonons {\em autonomously} drives the DQD charge qubit to a steady state that has coherence in the energy eigenbasis while retaining a significant degree of purity. This surprising result finds its explanation in the particular structure of the system-bath interaction~\cite{Giacomo_SSC}. Furthermore, the magnitude of steady-state coherence can be controlled by changing the experimentally tunable parameters of the qubit Hamiltonian, the detuning and the hopping. This is proven through an explicit calculation and characterization of the steady state Bloch vector of the charge qubit as a function of the controllable Hamiltonian parameters. { We also show that the coherence is also robust to presence of fermionic leads.}

In addition to the obvious importance of generating coherence for quantum information processing, there is currently significant interest in harnessing coherence and exploiting it a resource in other contexts~\cite{Streltsov2017}. In particular, coherence in the energy eigenbasis has been identified as one of the key features distinguishing quantum thermodynamics from its classical counterpart~\cite{Lostaglio2015nc,Lostaglio2015prx,narasimhachar2015low}. For example, coherence may enhance the performance of quantum refrigerators~\cite{Correa2014,Mitchison2015,Brask2015, Maslennikov2019} and heat engines~\cite{Scully2011,Ptaszynski2018,Klatzow2019} or be directly converted into work~\cite{allahverdyan2004maximal,korzekwa2016extraction,Kammerlander2016SciRep}. To address this point, we conclude the article by showing that, in our setup, the above phonon-induced steady-state coherence of the DQD can in fact be exploited to drive a mode of the surrounding cavity.

\section{Results}

\subsection{Autonomous generation of steady-state coherence} The main theoretical idea behind the autonomous generation of steady state coherence was first introduced and explored in~\cite{Giacomo_SSC}. In that work, sufficient conditions concerning the structure of the interaction Hamiltonian between a qubit and a bosonic bath were identified such to lead to steady-state coherence. In particular, it was shown that a spin-boson model with a Hamiltonian of the form
\begin{equation}\label{eq:Ham1}
\Ham = \frac{\omega_q}{2}\hat{\sigma}_z +\sum_k \Omega_k \hat{b}^{\dagger}_k \hat{b}_k + \left( f_1\hat{\sigma}_z + f_2\hat{\sigma}_x\right)\sum_{k} \lambda_k \left(\hat{b}_{k}^{\dagger} + \hat{b}_{k}\right)
\end{equation}
autonomously leads to a non-zero steady-state value for $\langle \hat{\sigma}_x\rangle$. Here, $\hat{\sigma}_{x,y,z}$ denote the usual Pauli spin operators and $f_{1}$, $f_{2}\neq 0$ two generic coupling constants, $\hat{b}_k$ is the bosonic annihilation operator of the $k$th bath mode. The results presented in this paper stem from the crucial observation that a semiconductor DQD in contact with a phononic substrate is described exactly by a Hamiltonian of the form of Eq.~\eqref{eq:Ham1}.

\begin{figure}
    \centering
    \includegraphics[width=\linewidth, trim = {7cm, 9cm, 7cm, 9cm}, clip]{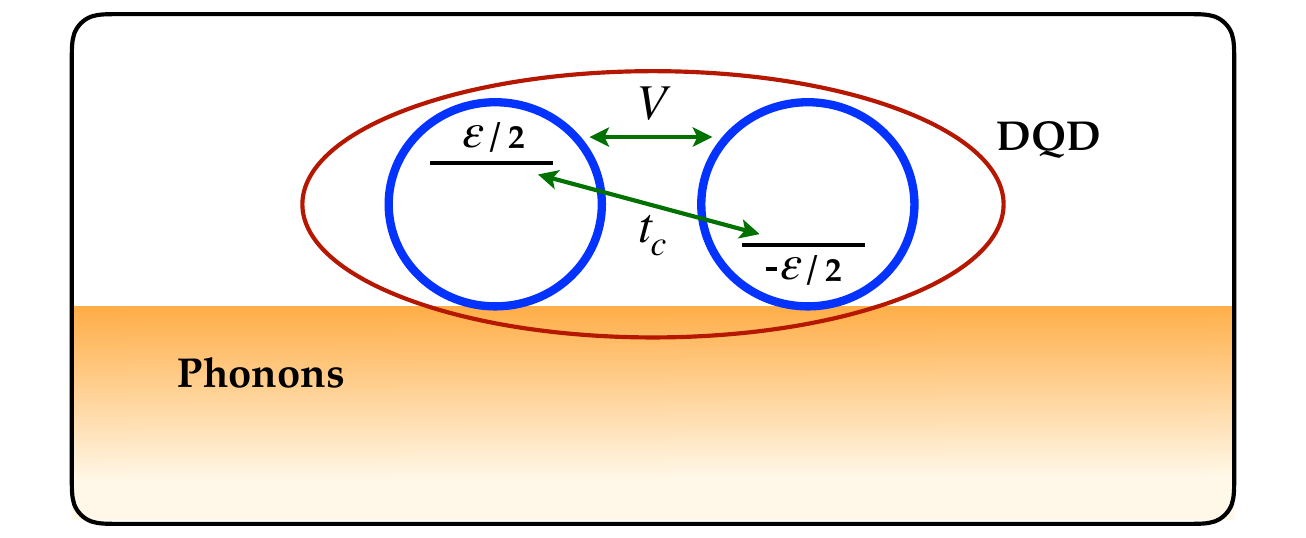}
    \caption{Schematic depiction of two quantum dots detuned in energy by $\varepsilon$ with inter-dot tunnelling $t_c$ and Coulomb repulsion $V$, interacting with a substrate supporting phononic excitations.}
    \label{fig:schematic}
\end{figure}

 \subsection{The DQD charge qubit} The set-up we consider is depicted schematically in Fig.~\ref{fig:schematic}. The DQD comprises two fermionic modes with strong repulsive interaction between them. Substrate phonons are coupled to the electric dipole moment of the DQD.  The full Hamiltonian (Refs.~\cite{Manas_photon_gain,Petta2, Petta4}) is then given by $\hat{\mathcal{H}}=\hat{\mathcal{H}}_{S}+\hat{\mathcal{H}}_{SE}+\hat{\mathcal{H}}_{E}$ with
\begin{align}
&\hat{\mathcal{H}}_{S} + \hat{\mathcal{H}}_{SE} = 
\left (\frac{\varepsilon}{2}+\hat{B}\right)\left(\hat{n}_1-\hat{n}_2\right)+ t_c \left(\hat{c}_1^\dagger\hat{c}_2 + \hat{c}_2^\dagger\hat{c}_1 \right)\notag \\ 
& \qquad \qquad \qquad + V\hat{n}_1\hat{n}_2, \nonumber \\
&\hat{B}=\sum_{k}\lambda_k \left(\hat{b}_{k}^{\dagger} + \hat{b}_{k}\right),\quad \hat{\mathcal{H}}_{E} = \sum_{k}\Omega_{k} \hat{b}_{k}^{\dagger} \hat{b}_{k}, \nonumber \\
&\hat{\mathcal{H}}_{S}=\frac{\varepsilon}{2}\left(\hat{n}_1-\hat{n}_2\right)+ t_c \left(\hat{c}_1^\dagger\hat{c}_2 + \hat{c}_2^\dagger\hat{c}_1 \right)+V\hat{n}_1\hat{n}_2,~~ \hat{\mathcal{H}}_{SE}=\hat{B}\left(\hat{n}_1-\hat{n}_2\right).
\end{align}
Here $\hat{n}_\ell=\hat{c}_\ell^\dagger \hat{c}_\ell$,   $\hat{c}_\ell$ is the fermionic annihilation operator of the $\ell$th site and $\hat{b}_{k}$ is the phononic annihilation operator of the $k$th mode of the bath. The experimentally controllable parameters of the DQD are the detuning $\varepsilon$ and the hopping $t_c$. The repulsive interaction between the two sites is given by $V$, which is usually much larger than any other energy scale in the regime of operation. The operator $\hat{B}$ embodies the noisy detuning due to the fluctuating phonon bath. Assuming that the latter is in a thermal state relative to an inverse temperature $\beta$, i.e. $\rho_E = Z_E^{-1} \exp(-\beta \hat{\mathcal{H}}_{E})$ ($Z_E = \mathrm{Tr}_E(\exp(-\beta \hat{\mathcal{H}}_{E}))$), the noise is characterized by zero mean value, i.e. $\langle \hat{B}(t) \rangle_E = 0 $, and the auto-correlation function
\begin{align}
& \langle\hat{B}(t)\hat{B}(0)\rangle_E=\int_0^{\infty}\!\!\!d\omega\left[W_s(\omega,\beta) \cos(\omega t) - i W_a(\omega) \sin(\omega t)\right] \nonumber \\
& W_s(\omega,\beta) = \mathfrak{J}_{ph}(\omega) \coth\left(\frac{\beta\omega}{2}\right),~~W_a(\omega) = \mathfrak{J}_{ph}(\omega),
\end{align}   
where $\mathfrak{J}_{ph}(\omega)= \sum_k \lambda_k^2 \delta(\omega-\Omega_k)$ is the spectral function of the phononic bath and $\langle ... \rangle_E= \Tr(\rho_E ..)$. In the above equation, $W_s(\omega,\beta) $ and $W_a(\omega) $ represent the symmetric and the anti-symmetric parts of the noise power spectral density, respectively. The presence of $W_a(\omega)$ is  the hallmark of `quantum noise'~\cite{quantum_noise}. Phenomenologically setting  
$W_a(\omega)=0$ models `classical noise'. This, for example, is approximately the case at high temperatures, where $W_a(\omega)$ is negligible compared to $W_s(\omega,\beta)$.

{
The system Hamiltonian $\hat{\mathcal{H}}_S$ can be diagonalized by transforming to the fermionic operators $\hat{A}_\alpha$, which are related to $\hat{c}_\ell$ via the following transformation,
\begin{align}
\label{def_theta}
&\left(
\begin{array}{c}
\hat{A}_1\\
\hat{A}_2\\
\end{array}
\right)= \Phi \left(
\begin{array}{c}
\hat{c}_1\\
\hat{c}_2\\
\end{array}
\right), ~~
\Phi=
\left(
\begin{array}{cc}
\cos(\frac{\theta}{2}) & \sin(\frac{\theta}{2})\\
-\sin(\frac{\theta}{2}) & \cos(\frac{\theta}{2})\\
\end{array}
\right), \nonumber \\
&\theta = \tan^{-1}\left(\frac{2 t_c}{\varepsilon}\right).
\end{align}
In the transformed basis, we have
\begin{align}
&\hat{\mathcal{H}}_{S} = \frac{\omega_q}{2} (\hat{N}_1 - \hat{N}_2) + V\hat{N}_1\hat{N}_2,~ \omega_q = \sqrt{\varepsilon^2 + 4 t_c^2},   \\
& \hat{\mathcal{H}}_{SE} = \Big[\cos\theta(\hat{N}_1-\hat{N}_2)- \sin\theta(\hat{A}_1^\dagger\hat{A}_2+\hat{A}_2^\dagger\hat{A}_1)\Big]\hat{B},  \nonumber \\
& \hat{N}_1=\hat{A}_1^\dagger\hat{A}_1,~~\hat{N}_2=\hat{A}_2^\dagger\hat{A}_2.
\end{align}
The DQD can be operated in the single-particle regime, i.e., where $\hat{N}_1+\hat{N}_2=1$.  In this regime, the fermionic operators can be exactly mapped to Pauli spin operators, 
\begin{align}
\hat{\sigma}_z=\hat{N}_1-\hat{N}_2, ~\hat{\sigma}_x=\hat{A}_1^\dagger\hat{A}_2+\hat{A}_2^\dagger\hat{A}_1,
\end{align}
and $\hat{\sigma}_y = i [\hat{\sigma}_x,\hat{\sigma}_z]/2=-i(\hat{A}_1^\dagger\hat{A}_2-\hat{A}_2^\dagger\hat{A}_1)$. The resulting Hamiltonians for the DQD and the DQD-phonon couplings become, in the eigenbasis of the DQD Hamiltonian,
\begin{align}
\label{def_spin_boson}
&\hat{\mathcal{H}}_{S} = \frac{\omega_q}{2}\hat{\sigma}_z , ~\hat{\mathcal{H}}_{SE}=\left( f_1\hat{\sigma}_z +f_2 \hat{\sigma}_x \right) \hat{B},
\end{align}
where $f_1=\cos\theta=\varepsilon/\omega_q$ and $f_2=-\sin\theta=-2t_c/\omega_q$.} The above equation describes a DQD charge qubit coupled to a phononic bath. Whatever state the charge qubit is prepared in, due to the phonons, the charge qubit relaxes to a unique steady state. To determine the latter, we notice that the above Hamiltonian shares exactly the same form as Eq.~\eqref{eq:Ham1}. This means that \emph{in a solid-state DQD charge qubit, the presence of phonons can autonomously generate coherence in the eigenbasis of the system Hamiltonian}. 

{
In general, the quantum dots experience correlated noise due to their coupling to a common phonon bath. However, this feature is not necessary for the mechanism of coherence generation considered here, in contrast to schemes for bath-induced entanglement production where environmental correlations are essential~\cite{Benatti2003,Solenov2007,McCutcheon2009,Zell2009}. Indeed, steady-state coherences would arise even if each quantum dot were coupled to its own independent bath, as shown in the Supplementary Notes. The essential ingredient for our study is the competition between the hopping between the dots and the local coupling of each quantum dot to the bath. This condition ensures that $\hat{\mathcal{H}}_{SE}$ contains both $\hat{\sigma}_z$ and $\hat{\sigma}_x$ components, as required for the autonomous generation of steady-state coherence~\cite{Giacomo_SSC}. We now proceed to characterize this coherence and investigate its engineering and tunability via the DQD parameters $\varepsilon$ and $t_C$.
}
\subsection{Steady-state properties}
The state of a qubit is completely characterized by the expectation values of the three Pauli operators $\hat{\sigma}_{x,y,z}$. In the long-time limit, the entire set-up reaches an equilibrium steady state where the charge current, which is proportional to $\langle \hat{\sigma}_y \rangle$, is zero. The remaining two non-zero expectation values are found to be (see Methods)
\begin{align}
\label{sigmaxz}
\langle\hat{\sigma}_x\rangle & = -\frac{\sin 2\theta}{\omega_q}\left[\Delta_s(\omega_q,\beta)\langle\hat{\sigma}_z\rangle_0  + \Delta_a(\omega_q)-\Delta_a(0)\right],  \\
\langle\hat{\sigma}_z\rangle & = \langle\hat{\sigma}_z\rangle_0 + \sin^2\theta\Big[\Delta_-(\omega_q)+\Delta_+(\omega_q) \nonumber \\
&\quad -\frac{\beta}{2}\textrm{sech}^2\left(\frac{\beta \omega_q}{2}\right) \Delta_s(\omega_q,\beta)\Big],  
\end{align}
where $\langle\hat{\sigma}_z\rangle_0 = -W_a(\omega_q)/W_s(\omega_q,\beta)=-\tanh\left(\beta\omega_q/2\right)$ and $\Delta_s(\omega_q,\beta)$, $\Delta_a(\omega_q)$ and $\Delta_{\pm}(\omega_q,\beta)$ are the principal-value integrals defined below:
\begin{align}
   \Delta_s(\omega_q,\beta) &=\mathcal{P} \int_0^\infty W_s(\omega,\beta)\left[\frac{1}{\omega+\omega_q}-\frac{1}{\omega-\omega_q} \right], \\
    \Delta_a(\omega_q) & =\mathcal{P} \int_0^\infty d\omega W_a(\omega)\left[\frac{1}{\omega+\omega_q}+\frac{1}{\omega-\omega_q} \right], \\ \Delta_{\pm}(\omega_q,\beta) & =\mathcal{P} \int_0^\infty d\omega\frac{\langle\hat{\sigma}_z\rangle_0 W_s(\omega,\beta)\pm W_a(\omega)}{(\omega\pm \omega_q)^2}.
\end{align}
These results show that if $\sin2\theta\neq 0$, which corresponds to $\varepsilon, t_c\neq 0$, and if the noise is quantum, i.e., $ W_a(\omega)\neq 0$,  we have $\langle\hat{\sigma}_x\rangle\neq 0$. Thus, in this case, there will be coherence in the eigenbasis of the system in the steady state. If, on the other hand, the noise were classical, i.e., $ W_a(\omega)= 0$, we would obtain $\langle\hat{\sigma}_x\rangle= 0$ and there would be no steady state coherence. This implies that, in the high-temperature regime that corresponds to the classical limit~\cite{quantum_noise}, no steady-state-coherence will be generated. This can also be checked by taking $\beta\rightarrow 0$ limit of the above results. In the opposite limit of low temperatures,  $\beta\omega_q\gg 1$, the above formulae simplify to
\begin{align}
\label{sigmax_zero_temp}
&\langle\hat{\sigma}_x\rangle =-2\sin 2\theta\int_0^\infty  d\omega\frac{\mathfrak{J}_{ph}(\omega)}{\omega(\omega+\omega_q)}  \\
\label{sigmaz_zero_temp}
&\langle\hat{\sigma}_z\rangle = -1 + 2\sin^2\theta\int_0^\infty d\omega \frac{\mathfrak{J}_{ph}(\omega)}{(\omega+\omega_q)^2}.
\end{align}
The above equations show that, within this temperature regime, the state of the DQD becomes temperature-independent. Note, however, that these expressions are only valid for temperatures well above the Kondo scale $T_K$, where our perturbative analysis is expected to break down \cite{spin_boson_Kondo1,spin_boson_Kondo2,spin_boson_Kondo3}. Nevertheless, $T_K$ is exponentially suppressed by weak system-bath coupling, thus providing a wide temperature regime where our results hold.  

\subsection{Numerical results} 
\begin{figure}
\includegraphics[width=\columnwidth]{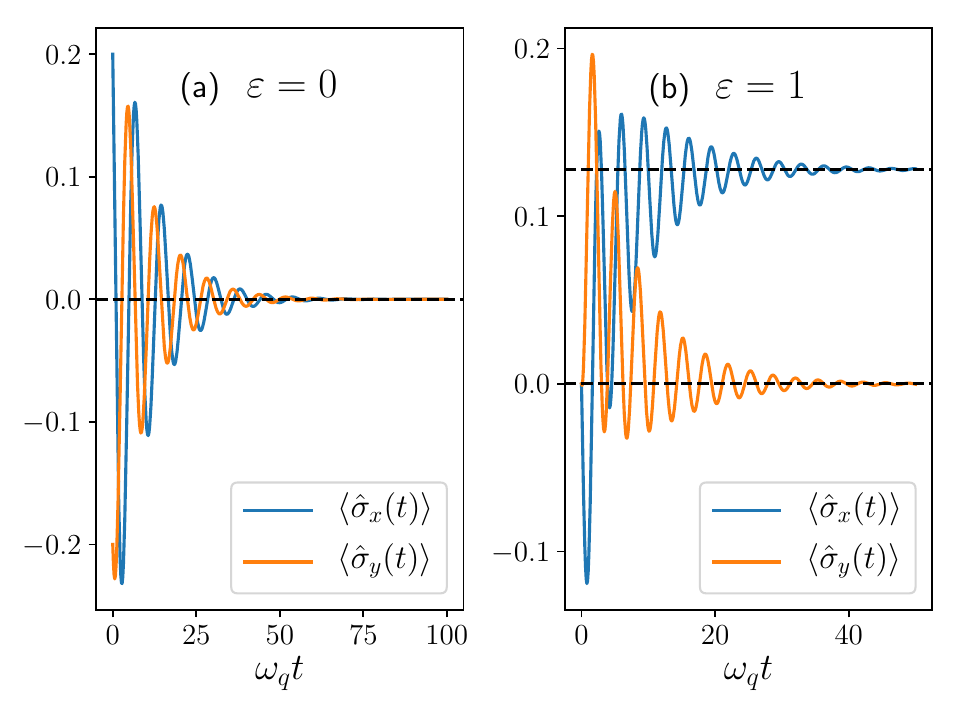} 
\caption{ The figure shows decoherence and coherence generation in DQD charge qubit due to presence of a phononic environment. (a) Decoherence: Dynamics of $\langle\hat{\sigma}_x(t)\rangle$ and $\langle\hat{\sigma}_y(t)\rangle$ for $\varepsilon=0$, starting from an initial state with coherence $\langle \hat{\sigma}_x (0)\rangle=0.2$, $\langle \hat{\sigma}_y (0)\rangle=-0.2$, $\langle \hat{\sigma}_z (0) \rangle=0$. Here the final steady state has no coherence. (b) Coherence generation: Dynamics of $\langle\hat{\sigma}_x(t)\rangle$ and $\langle\hat{\sigma}_y(t)\rangle$ for $\varepsilon=1$, starting from a completely mixed initial state: $\langle \hat{\sigma}_x (0) \rangle=0$, $\langle \hat{\sigma}_y (0)\rangle=0$, $\langle \hat{\sigma}_z (0)\rangle=0$.  Here the final steady state has coherence. Other parameters: $t_c=0.5$, $\gamma_b=0.03$, $\omega_{max}=10$. All energies are measured in units of $\omega_c$, which is set to $1$.}
\label{fig:dynamics} 
\end{figure}

Above, we have found that quantum noise due to phonons can be used to generate coherence in the energy eigenbasis of the DQD charge qubit for non-zero detuning.  This is shown in Fig.~\ref{fig:dynamics}, where the dynamics of $\langle\hat{\sigma}_x(t)\rangle$ and $\langle\hat{\sigma}_y(t)\rangle$ are plotted for two different cases showing the effects of decoherence (for $\varepsilon=0$) and coherence generation (for $\varepsilon=1)$ (see Methods). 
In our calculations, we have taken the phonon spectral function as
\begin{align}\label{spectral_function}
\mathfrak{J}_{ph}(\omega)=\gamma_b\omega\left[1-\textrm{sinc}\left(\frac{\omega}{\omega_{c}}\right)\right]e^{-{\omega^2}/{2\omega_{max}^2}}.
\end{align}
The spectral functions of phonons in solid-state DQDs have been well characterized theoretically and experimentally and they vary depending on the experimental platform~\cite{Petta1, Petta2, phonon_spectral_deriv, measuring_phonons_2010,Colless2014}. Our chosen spectral function is known to be a good description of bulk acoustic phonons in GaAs DQDs~\cite{Petta2, measuring_phonons_2010,Colless2014}. The frequency $\omega_{max}$ is the upper cut-off frequency, while $\omega_c=c_s/d$, where $c_s$ is the speed of sound in the substrate and $d$ is the distance between the two quantum dots.  The dimensionless parameter $\gamma_b$ controls the strength of coupling with phonons. The validity of our theory requires that $\gamma_b\ll 1$.  We have set $\omega_c=1$ and used this as our unit of energy.

\begin{figure}
\includegraphics[width=\columnwidth]{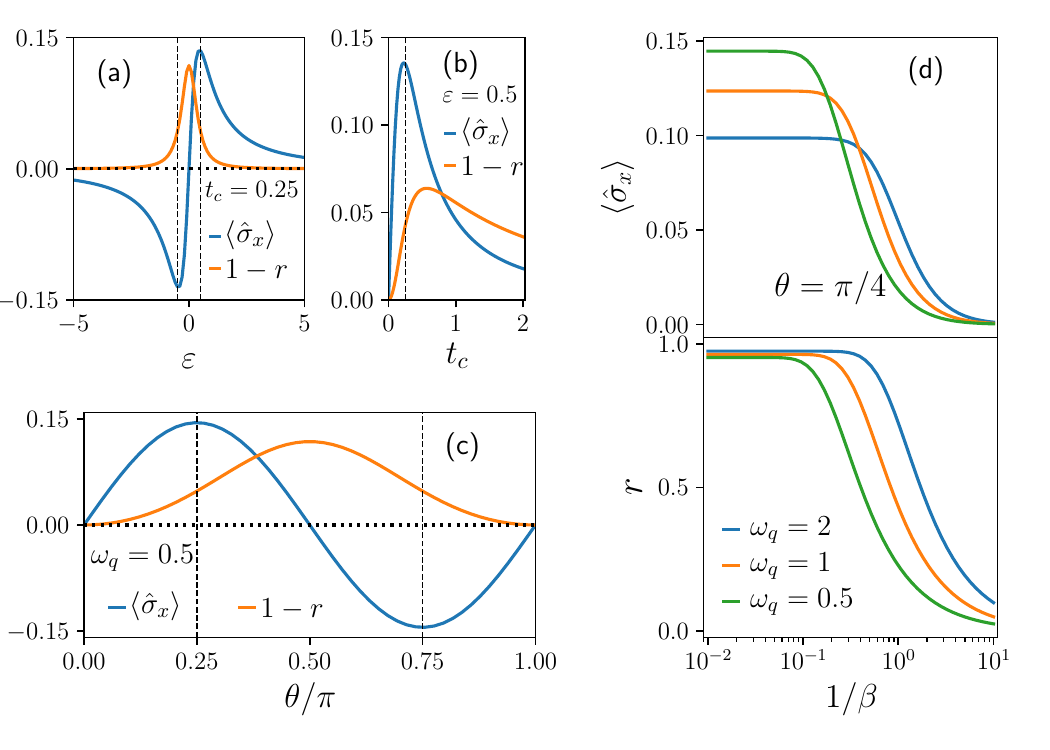} 
\caption{ The figure shows the variation of $\langle\hat{\sigma}_x\rangle$ and length of Bloch vector $r$  with several parameters of DQD charge qubit. (a) $\langle\hat{\sigma}_x\rangle$ and $1-r$ vs the detuning $\varepsilon$ at fixed hopping $t_c=0.25$ at low temperature, (b) $\langle\hat{\sigma}_x\rangle$ and $1-r$ vs the hopping $t_c$ at fixed detuning $\varepsilon=0.5$ at zero temperature, (c) $\langle\hat{\sigma}_x\rangle$ and $1-r$ vs $\theta$ at fixed $\omega_q$ at zero temperature. The vertical dashed lines in these plots correspond to $\varepsilon=\pm 2t_c$. (d)   $\langle\hat{\sigma}_x\rangle$ (top) and $r$ (bottom) vs temperature $1/\beta$ at fixed $\theta=\pi/4$ and for various chosen values of $\omega_q$. Other parameters: $\gamma_b=0.03$, $\omega_{max}=10$. All energies are measured in units of $\omega_c$, which is set to $1$.  $\langle\hat{\sigma}_y\rangle=0$.}
\label{fig:qubit_state} 
\end{figure}

In Fig.~\ref{fig:qubit_state} we characterize the steady state of the qubit. Since $\langle\hat\sigma_y\rangle=0$, we parameterize the steady state with $\langle\hat{\sigma}_x\rangle$ and the length of the Bloch vector
\begin{align}
r=\sqrt{\langle\sigma_x\rangle^2+\langle\sigma_z\rangle^2}.
\end{align}
While $\langle\hat{\sigma}_x\rangle$ is a measure of coherence, $r$ is a measure of purity of the state. For a pure state, $r=1$, for a completely mixed state, $r=0$. Fig.~\ref{fig:qubit_state} demonstrates that tunable steady-state coherence persists with high purity across a range of temperatures and energy scales. The maximal value of $\langle \hat{\sigma}_x\rangle$ can be achieved at low temperatures by setting the dots' detuning to $\varepsilon = \pm 2 t_c$. The sign of the coherence can be made positive or negative, depending on the sign of $\sin 2\theta = 4\varepsilon t_c/\omega_q^2$, as can also be seen directly from Eq.~\eqref{sigmaxz}. Interestingly, the coherence and purity remain constant as the temperature grows up to values of order $k_B T \sim 0.1 \omega_q$, and then decay to zero as the temperature is further increased. It should therefore be possible to observe maximal phonon-induced steady-state coherence in a DQD system at only moderately low temperatures. 

Let us make an order-of-magnitude estimate in order to demonstrate the feasibility of our proposal. We take $\omega_q=0.5\omega_c$, $\varepsilon=2t_c$ and $\beta\omega_q=10$, corresponding to the maximum coherence in Fig.~\ref{fig:qubit_state}. In GaAs, $c_s\approx 3000$m/s, so that an inter-dot separation of $d = 150$nm yields $\omega_c\approx 20$GHz. Therefore, the maximal coherence is obtained when $\varepsilon\approx 29\mu$eV, $t_c\approx 15\mu$eV and $T \approx 50$mK.  These parameters are completely within reach of current experiments  (for example, see Refs.~\cite{Petta1,Petta2}). 

{
\subsection{Robustness to the presence of fermionic leads}

Experimental set-ups such as Refs.~\cite{Petta1,Petta2} typically feature fermionic leads coupled to the DQD. In this case, even when the leads are in equilibrium, the number of particles in the DQD is not strictly conserved. However, the DQD can only be considered as a qubit in the single-particle sector. Nevertheless, we now show that steady-state coherence is robust to the presence of fermionic leads at equilibrium, i.e, at equal chemical potential $\mu$, and at the same temperature at the substrate phonons. 

The total Hamiltonian of the set-up is now given by $\hat{\mathcal{H}}=\hat{\mathcal{H}}_{S}+\hat{\mathcal{H}}_{SE}+\hat{\mathcal{H}}_{E}+\hat{\mathcal{H}}_{Sf}+\hat{\mathcal{H}}_{f}$,
$
\hat{\mathcal{H}}_{Sf} = \sum_{\ell=1,2}\sum_{r=1}^{\infty} \gamma_{r\ell} \left(\hat{c}_\ell^\dagger\hat{B}_{\ell r} + \hat{B}_{\ell r}^\dagger \hat{c}_\ell\right),~\hat{\mathcal{H}}_{f} = \sum_{\ell=1,2}\sum_{r=1}^{\infty}\mathcal{E}_{\ell r} \hat{B}_{\ell r}^\dagger \hat{B}_{\ell r},  
$ 
where the fermionic lead is modelled by infinite number of fermionic modes, and $\hat{B}_{\ell r}$ is the annihilation operator for the $r$th mode of the fermionic lead attached to the $\ell$th DQD site. For simplicity, we assume that the leads have identical, constant spectral functions $\mathfrak{J}^f_\ell(\omega) = 2\pi \sum_{r=1}^\infty \gamma_{\ell r}^2 \delta(\omega-\mathcal{E}_{\ell r}),$ $\mathfrak{J}^f_1(\omega)= \mathfrak{J}^f_2(\omega)=\Gamma$, for $-\Lambda\leq \Gamma \leq \Lambda$, and zero otherwise.
Here, we assume that the coupling to the leads $\Gamma$ is weak, while the bandwidth $\Lambda$ is assumed to be large. 

\begin{figure}
\includegraphics[width=\columnwidth]{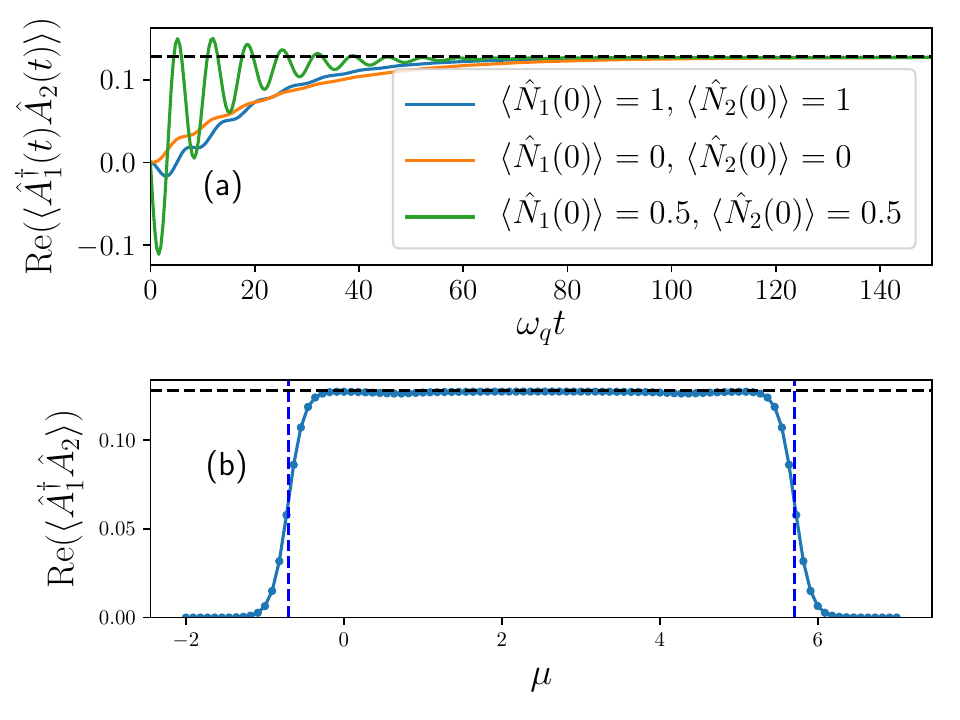} 
\caption{{  The figure shows the robustness of phonon-induced steady state coherence in a DQD charge qubit to presence of fermionic leads at equilibrium.  (a) Evolution of $\textrm{Re}\Big(\langle\hat{A}_1^\dagger(t)\hat{A}_2(t)\rangle \Big)$ in presence of the fermionic lead is shown for three different initial conditions of the DQD with no initial coherence. The chemical potential of the fermionic lead is $\mu=0$.  (b) The equilibrium steady state value of $\textrm{Re}\Big(\langle\hat{A}_1^\dagger\hat{A}_2\rangle \Big)$ as a function of $\mu$.      The vertical dashed lines show positions of $\mu=-\omega_q/2$ and $\mu=-\omega_q/2+V$. The horizontal dashed lines in both (a) and (b) show the value of coherence given by the expression for $\langle\hat{ \sigma}_x \rangle$ in Eq.~\eqref{sigmax_zero_temp}. Parameters for both plots: $\varepsilon=1$, $t_c=0.5$, $\mu=0$, $V=5$, $\beta=10$, $\gamma_b=0.03$, $\omega_{max}=10$, $\Gamma=0.06$, $\Lambda=400$.  All energies are measured in units of $\omega_c$, which is set to $1$.}}
\label{fig:with_leads} 
\end{figure}

At low temperatures, the number of particles in the DQD in the equilibrium steady state is governed by the chemical potential of the fermionic lead. The DQD is occupied by a single particle on average if the chemical potential is in the regime 
\begin{align}
\label{chemical_pot_condition}
-\frac{\omega_q}{2} \ll \mu \ll \frac{\omega_q}{2}+ V,
\end{align}
i.e., the chemical potential is higher than $\omega_q/2$, but smaller than the charging energy of the DQD. Therefore, when Eq.~\eqref{chemical_pot_condition} holds, we expect the system to display steady-state coherence if there is non-zero detuning.

To check this, we calculate $\textrm{Re}\Big(\langle\hat{A}_1^\dagger(t)\hat{A}_2(t)\rangle \Big)$, which corresponds to $\langle\hat{\sigma}_x(t)\rangle$ in the single-particle sector (see Methods). The dynamics of this quantity for three different initial conditions of the DQD are shown in Fig.~\ref{fig:with_leads}(a). The three initial conditions correspond to the DQD being doubly occupied, the DQD being completely unoccupied, and the completely mixed state of the charge qubit, which corresponds to $\langle \hat{N}_1(0) \rangle=0.5$, $\langle \hat{N}_2(0) \rangle=0.5$. None of the initial states contain any coherence in the energy eigenbasis. The chemical potential of the fermionic lead is chosen as $\mu=0$, which satisfies Eq.~\eqref{chemical_pot_condition}. For non-zero detuning, $\textrm{Re}\left(\langle\hat{A}_1^\dagger(t)\hat{A}_2(t)\rangle \right)$ goes to the same steady-state value given by the expression for  $\langle\hat{\sigma}_x\rangle$ in Eq.~\ref{sigmax_zero_temp} for all initial conditions. Fig.~\ref{fig:with_leads}(b) shows the plot of the equilibrium steady state value of $\textrm{Re}\left(\langle\hat{A}_1^\dagger\hat{A}_2\rangle \right)$ as a function of $\mu$. In the regime corresponding to Eq.~\eqref{chemical_pot_condition}, there is coherence in the energy eigenbasis, while beyond that regime, the coherence decays to zero. Thus, we have shown that the coherence generated due to the presence of phonons is completely robust to presence of equilibrium fermionic leads, so long as the DQD is in the single-particle sector.
}

\subsection{Application to driving a cavity}

As an application, we now demonstrate that phonon-induced coherence may be used to displace an auxiliary cavity mode. Let the charge qubit be coupled to a cavity mode described by a Jaynes-Cummings like interaction
\begin{align}
\hat{\mathcal{H}}_{C}= \omega_0 \hat{a}^\dagger\hat{a},~~\hat{\mathcal{H}}_{SC}=g(\hat{\sigma}_+ \hat{a} +  \hat{a}^\dagger\hat{\sigma}_-).
\end{align}
Here $\omega_0$ is frequency of the cavity mode, $\hat{a}$ is bosonic annihilation operator for the cavity mode, $g$ is the cavity-DQD coupling strength and $\hat{\sigma}_{\pm}=(\hat{\sigma}_x\pm i\hat{\sigma}_y)/2$. The cavity will be further coupled with its own thermal environment, leading to a decay rate $\kappa$.

\begin{figure}
\includegraphics[width=\columnwidth]{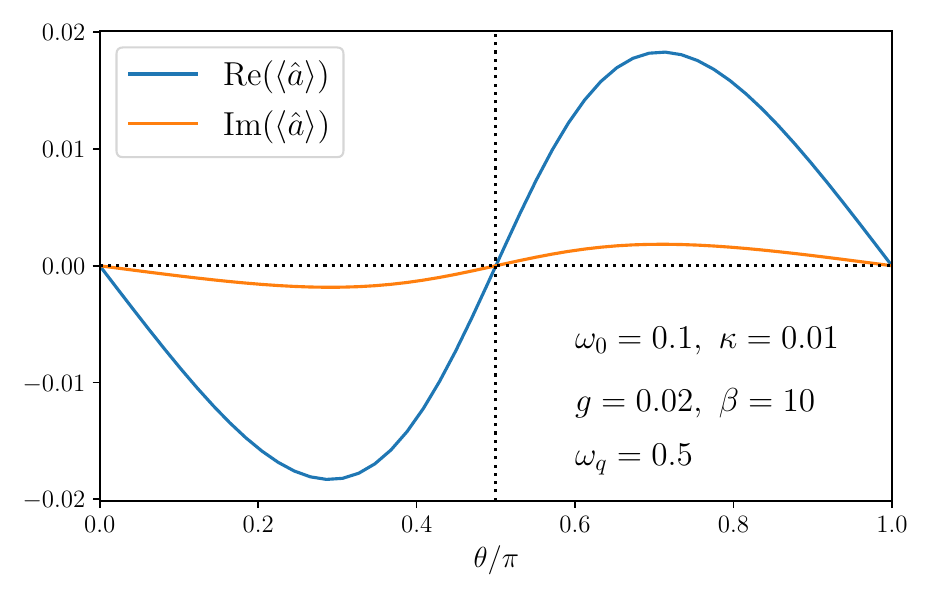} 
\caption{ A cavity with a weak Jaynes-Cummings coupling to the DQD will be displaced due to the steady state coherence of the DQD. The figure shows plots of real and imaginary parts of $\langle \hat{a} \rangle$, as a function of DQD parameter $\theta$ with other parameters fixed.  Other parameters: $\gamma_b=0.03$, $\omega_{max}=10$. All energies are measured in units of $\omega_c$, which is set to $1$.}
\label{fig:cavity_state} 
\end{figure}

If both the cavity-DQD coupling and the cavity decay rate are small, and if the cavity is away from resonance with the DQD (i.e, $\omega_q\neq \omega_0$), then we can obtain the steady state results for the cavity up to leading order in small quantities. The results for the expectation value of the cavity field operator $\langle \hat{a} \rangle$ and the cavity occupation $\langle\hat{a}^\dagger \hat{a} \rangle$ to leading order are given by
\begin{align}
\langle\hat{a}\rangle= -\frac{ g}{2(\omega_0-i\kappa)}\langle  \hat{\sigma}_x\rangle,~\langle \hat{a}^\dagger \hat{a} \rangle\simeq\frac{1}{e^{\beta\omega_0}-1} + O(g^2),
\end{align}
while that for $\langle\hat{a}^2\rangle=0$. Thus, it is immediately clear that \emph{the coherence of the charge qubit  causes displacement of the cavity mode}, given by $\langle\hat{X}\rangle=\langle\hat{a}+\hat{a}^\dagger\rangle\simeq -2g\langle  \hat{\sigma}_x\rangle/\omega_0$. However, $\langle \hat{a}^\dagger \hat{a} \rangle$ is given by the Bose-Einstein distribution for the cavity, which ensures that no average photon current flows between the cavity and its own bosonic thermal bath. This is consistent with the fact that entire set-up is at thermal equilibrium.  This means that the variance of $\hat{X}$ is given by $\textrm{var}(\hat{X})=\langle \hat{X}^2 \rangle - \langle \hat{X} \rangle^2\simeq \coth(\beta\omega_0/2) + O(g^2)$.   The cavity displacement $\langle\hat{X}\rangle$, which may in principle be experimentally measured \cite{light_measurement,Bakker2015}, thus directly probes the steady-state coherence of the DQD.
In Fig.~\ref{fig:cavity_state}, we show the real and imaginary parts of $\langle \hat{a} \rangle$ in steady state as the DQD parameter $\theta$ is tuned. It is clear that $\langle \hat{a} \rangle$ is non-zero when $\langle\hat{\sigma}_x\rangle$ is non-zero.

\section{Discussion}

We have theoretically demonstrated that quantum noise due to phonons in state-of-the-art DQD charge qubits autonomously generates steady-state coherence in the energy eigenbasis of the qubit at non-zero detuning. Due to the inherently dissipative nature of this phenomenon, it represents an especially robust yet surprisingly tunable addition to the toolbox of quantum state engineering for solid-state charge qubits, and promises potential to become a new resource for quantum thermodynamics. Remarkably, the magnitude and sign of the coherence can be controlled merely by manipulating the Hamiltonian parameters of the DQD, while retaining high purity. Aside from the intrinsic interest of generating quantum coherence, the resulting steady states represent useful resources in various contexts. As a simple example, we have shown that the qubit coherence in turn generates field coherence in a cavity mode that is weakly coupled to the DQD. { Moreover, coherence distillation or amplification techniques could in principle be applied to generate fully coherent resource states for quantum information processing~\cite{Winter2016,Liu2019,Manzano2019}.

We emphasize that the physics described here is a general property of spin-boson models of the type in Eq.~\eqref{def_spin_boson}. Such models can generically be used to describe dissipative two-level systems~\cite{Leggett1987,Breuer2002,thorwart2004dynamics}, and can be engineered in various platforms (for examples, see Refs.~\cite{KLHur_spin_boson1,qubit_noise_review} and citations therein). Indeed, any generic qubit Hamiltonian, $\hat{H} = \frac{\varepsilon}{2} \hat{\tau}_z +t_c \hat{\tau}_x$ with quantum noise in the qubit parameters $\varepsilon$ and $t_c$ can be written in the form of Eq.~\eqref{def_spin_boson} (see Supplementary Notes for details). It follows that, at low temperatures, if $f_1 f_2 \neq 0$, which is generically true, quantum noise in qubit parameters leads to generation of steady-state coherence in the energy eigenbasis of \textit{any} qubit, e.g. a superconducting qubit, a DQD spin qubit etc. On the other hand, classical noise in qubit parameters  is detrimental to such coherence. The DQD charge qubit provides an especially interesting example, since the required parameter regime corresponds to current state-of-the-art experiments \cite{Petta1,Petta2,Petta3,Petta4,Petta5}, where quantum noise in detuning is the major source of noise. 
}

Future work will focus on a full characterization of thermodynamic properties of the setup including an analysis of the relation between irreversible entropy production and coherence \cite{santos2019role,francica2019role}, which will shed light on the thermodynamic cost of generating steady-state coherence.

\section{Methods}
{

Detailed methods are given in the Supplementary Notes. Here we give a brief overview of the techniques.

\subsection{Redfield master equation}

The time evolution of the DQD qubit is modelled using a Redfield master equation, detailed in the Supplementary Notes. The derivation of this equation starts from a factorised initial system-environment state $\rho_{\rm tot}(0) = \rho(0)\otimes \rho_E$, with $\rho_E$ the thermal equilibrium state of the phononic bath and $\rho(0)$ the initial DQD state. The Redfield equation is based on two approximations: (i) a perturbative expansion up to second order in the coupling between the DQD and the reservoir, and (ii) a Markov assumption that the reservoir memory time is short in comparison to the time scale of the DQD evolution. In the presence of fermionic leads, the same set of approximations is used but where the leads are also incorporated into the environment. The result is a master equation describing the DQD of the form
\begin{equation}
\label{redfield_main}
    \frac{\partial\rho}{\partial t} = i[\rho,\hat{\mathcal{H}}_S] - \mathcal{L}_{ph}\rho - \mathcal{L}_{f}\rho,
\end{equation}
where $\mathcal{L}_{ph}$ and $\mathcal{L}_{f}$ respectively describe the effect of the phononic and fermionic reservoirs (see Supplementary Notes). The solution of Eq.~\eqref{redfield_main} is used to compute the time evolution of expectation values in the main text.

\subsection{Perturbation expansion of the generalized equilibrium state}

As discussed in Refs.~\cite{accuracy_of_qme}, the predictions of the Redfield equation give results for the DQD coherence valid up to second order in $\hat{\mathcal{H}}_{SE}$, while the predictions for the populations are valid only to zeroth order. However, in order to discuss the purity of the qubit, we need results for both coherences and populations valid to the same order. In order to achieve this, we use the fact that an open quantum system coupled weakly to a thermal environment is generically expected to relax to the generalized equilibrium state~\cite{Dhar2012,Juzar2012,Juzar2013,Juzar2017,Subasi2012PRE,alipour2019correlation}
\begin{equation}
    \label{generalised_equilibrium}
    \lim_{t\to \infty}\rho(t) = \Tr_E\left [\frac{{\rm e}^{-\beta ( \hat{\mathcal{H}}_S +  \hat{\mathcal{H}}_E +  \hat{\mathcal{H}}_{SE}) }}{\Tr\left [{\rm e}^{-\beta ( \hat{\mathcal{H}}_S +  \hat{\mathcal{H}}_E +  \hat{\mathcal{H}}_{SE}) }\right ]}\right ],
\end{equation}
which incorporates the effect of system-environment correlations. We use a perturbative expansion of Eq.~\eqref{generalised_equilibrium} up to second order in $\Ham_{SE}$ to obtain steady-state expectation values, as detailed in the Supplementary Notes. This approach yields an identical prediction for the coherence $\langle \hat{\sigma}_x\rangle$ as the master equation, as well as a second-order correction to the population inversion $\langle \hat{\sigma}_z\rangle$.
}

\section{Acknowledgments}
 We acknowledge funding from European Research Council Starting Grant ODYSSEY (Grant Agreement No. 758403). J.G. also acknowledges funding from a SFI Royal Society University Research Fellowship. R.F. acknowledges grant 19-17765S of Czech Science Foundation and national funding from the MEYS as well as funding from the European Union’s Horizon 2020 (2014–2020) research and innovation framework programme under grant agreement No. 731473 (project 8C18003 TheBlinQC). Project TheBlinCQ has received funding from the QuantERA ERA-NET CoFund in Quantum Technologies implemented within the European Union’s Horizon 2020 programme. 
\bibliography{ref_DQD_SSC}

\widetext
\clearpage
\begin{center}
\textbf{\large Supplemental Material: Tunable phonon-induced steady-state coherence in a double quantum dot}\\
\end{center}
\setcounter{equation}{0}
\setcounter{figure}{0}
\setcounter{table}{0}
\setcounter{page}{1}
\makeatletter
\renewcommand{\theequation}{S\arabic{equation}}
\renewcommand{\thefigure}{S\arabic{figure}}

In the Supplementary Notes, we provide the details of derivations of our analytical results in the main manuscript, as well as provide self-consistent tests for validity of our results. In Sec.~\ref{Sec:Fermion_to_qubit}, for completeness of the Supplementary notes, we provide to steps in going from fermionic language to the qubit language for the DQD. In Sec.~\ref{Sec:Independent_baths} we show that the same effective model of the DQD charge qubit is obtained if each fermionic site is coupled to its own independent phononic bath. In Sec.~\ref{Sec:DQD_RQME}, we give the derivation of our results via microscopically obtained quantum master equation in presence of only phonons (i.e, no fermionic leads). In Sec.~\ref{Sec:DQD_global_thermal_state}, we derive our results via perturbative expansion of a global thermal state. In Sec.~\ref{Sec:not_renormalized_qubit}, we show that the effects we discuss are completely distinguishable from a simple renormalization of qubit parameters due to presence of the phonoic environment. In Sec.~\ref{Sec:generality_of_results}, we show that quantum noise in qubit parameters of any qubit generically generates steady state coherence in the eigenbasis of the qubit. In Sec.~\ref{Sec:DQD_with_leads}, we derive the quantum master equation for the DQD in presence of the phononic bath as well as fermionic leads. In Sec.~\ref{Sec:Validity_of_RQME}, we mention some technical points regarding validity of results obtained from such a quantum master equation. In Sec,~\ref{Sec:cavity_calculations}, we give the derivation of results for the cavity with is coupled to the DQD charge qubit. 
 
\section{From fermions to qubit}\label{Sec:Fermion_to_qubit}

The full Hamiltonian of the set-up consisting of the DQD and the phononic bath is given by
\begin{align}
\label{H_set_up}
&\hat{\mathcal{H}}_{S} = \frac{\varepsilon}{2}\left(\hat{n}_1-\hat{n}_2\right)+V\hat{n}_1\hat{n}_2 + t_c \left(\hat{c}_1^\dagger\hat{c}_2 + \hat{c}_2^\dagger\hat{c}_1 \right), ~~\hat{\mathcal{H}}_{SE} = \left( \hat{n}_1 - \hat{n}_2 \right) \sum_{k=1}^\infty \lambda_k \left(\hat{b}_{k}^{\dagger} + \hat{b}_{k}\right), \\
& \hat{\mathcal{H}}_{E} = \sum_{k=1}^{\infty}\Omega_{k} \hat{b}_{k}^{\dagger} \hat{b}_{k},~~\hat{\mathcal{H}}=\hat{\mathcal{H}}_{S}+\hat{\mathcal{H}}_{SE}+\hat{\mathcal{H}}_{E}. \nonumber 
\end{align}
Here $\hat{n}_\ell=\hat{c}_\ell^\dagger \hat{c}_\ell$, and $\hat{c}_\ell$ is the fermionic annihilation operator of the $\ell$th site and $\hat{b}_{k}$ is the phononic annihilation operator of the $k$th mode of the bath. The DQD Hamiltonian can be diagonalized by the transforming to the fermionic operators $\hat{A}_\alpha$, which are related to $\hat{c}_\ell$ via the following transformaiton,
\begin{align}
\label{def_theta}
&\left(
\begin{array}{c}
\hat{A}_1\\
\hat{A}_2\\
\end{array}
\right)= \Phi \left(
\begin{array}{c}
\hat{c}_1\\
\hat{c}_2\\
\end{array}
\right), ~~
\Phi=
\left(
\begin{array}{cc}
\cos(\frac{\theta}{2}) & \sin(\frac{\theta}{2})\\
-\sin(\frac{\theta}{2}) & \cos(\frac{\theta}{2})\\
\end{array}
\right), ~~\theta = \tan^{-1}\left(\frac{2 t_c}{\varepsilon}\right).
\end{align}
In the transformed basis, the system Hamiltonian is given by,
\begin{align}
\label{def_omegaq}
&\hat{\mathcal{H}}_{S} = \frac{\omega_q}{2} (\hat{N}_1 - \hat{N}_2) + V\hat{N}_1\hat{N}_2, ~~\omega_q = \sqrt{\varepsilon^2 + 4 t_c^2},
\end{align}
where $\hat{N}_\alpha=\hat{A}_\alpha^\dagger \hat{A}_\alpha$.
In the eigenbasis of the DQD, the DQD-phonons coupling is given by
\begin{align}
&\hat{\mathcal{H}}_{SE} = \frac{1}{\omega_q}\Big[\varepsilon(\hat{N}_1-\hat{N}_2)- 2t_c(\hat{A}_1^\dagger\hat{A}_2+\hat{A}_2^\dagger\hat{A}_1)\Big] \sum_{s=1}^\infty \lambda_k \left(\hat{b}_{k}^{\dagger} + \hat{b}_{k}\right),
\end{align}
where $\hat{N}_\ell=\hat{A}_\ell^\dagger\hat{A}_\ell$.
Note that coupling with the phononic bath $\hat{\mathcal{H}}_{SE}$ is such that the total number of fermions is conserved, 
\begin{align}
&\hat{N}=\hat{n}_1+\hat{n}_2=\hat{N}_1+\hat{N}_2,~~[\hat{N}, \hat{\mathcal{H}}_{S}+\hat{\mathcal{H}}_{SE}+\hat{\mathcal{H}}_{E}]=0.
\end{align}
Since$\hat{N}_1\hat{N}_2 = \frac{\hat{N}^2-\hat{N}}{2}$,
$\hat{N}_1\hat{N}_2$ is also a conserved quantity.
We will assume that the initial state $\rho^{DQD}_{tot}$ is such that there is only one fermion in the DQD,
\begin{align}
\hat{N}=1,~~\hat{N}_1\hat{N}_2 =0
\end{align}
With this constraint, we exactly have
\begin{align}
\hat{\sigma}_z=\hat{N}_1-\hat{N}_2,~~ \hat{\sigma}_x=\hat{A}_1^\dagger\hat{A}_2+\hat{A}_2^\dagger\hat{A}_1,
\end{align}
where $\hat{\sigma}_z$ and $\hat{\sigma}_x$ are the usual Pauli spin operators.
So, we can exactly transform $\hat{\mathcal{H}}_{S}$ and $\hat{\mathcal{H}}_{SE}$ to
\begin{align}
\label{charge_qubit}
&\hat{\mathcal{H}}_{S} = \frac{\omega_q}{2}\hat{\sigma}_z, ~  \omega_q = \sqrt{\varepsilon^2 + 4 t_c^2},~~\hat{\mathcal{H}}_{SE} = \Big[ f_1 \hat{\sigma}_z +f_2 \hat{\sigma}_x\Big] \sum_{s=1}^\infty \lambda_k \left(\hat{b}_{k}^{\dagger} + \hat{b}_{k}\right),~~f_1 =\frac{\varepsilon}{\omega_q}\equiv \cos(\theta),~ f_2=-\frac{2t_c}{\omega_q} =- \sin(\theta).
\end{align} 
This is the Hamiltonian of the charge qubit. 

\section{Independent baths}\label{Sec:Independent_baths} 

We now show that effectively the same model is obtained if the two quantum dots are coupled to \textit{independent} baths. In order to illustrate the principle, it is sufficient to assume that the two baths have identical dispersion relations and couplings to their respective quantum dots. Consider in particular the total Hamiltonian $\hat{\mathcal{H}}'=\hat{\mathcal{H}}_{S}+\hat{\mathcal{H}}'_{SE}+\hat{\mathcal{H}}'_{E}$, where  $\hat{\mathcal{H}}'_{SE} = \sum_{j=1}^2 \hat{n}_j\hat{B}_j$ and $\hat{\mathcal{H}}'_E = \sum_{j=1}^2 \hat{\mathcal{H}}_{E,j}$, with
\begin{align}
\hat{B}_j & =  \sum_{k} \lambda'_{k} \left(\hat{b}_{k,j}^{\dagger} + \hat{b}_{k,j}\right),\\
\hat{\mathcal{H}}_{E,j} & =  \sum_{k} \Omega_k \hat{b}_{k,j}^{\dagger} \hat{b}_{k,j},
\end{align}
while $\hat{\mathcal{H}}_S$ is given in Eq.~\eqref{H_set_up}. The canonical operators $\hat{b}_{k,j}$ describe independent modes in each bath, i.e. $[\hat{b}_{k,j}, \hat{b}^\dagger_{k',j'}] = \delta_{kk'}\delta_{jj'}$. It is straightforward to rewrite the interaction Hamiltonian as follows
\begin{align}
\hat{\mathcal{H}}_{SE} & = \frac{1}{2} \left(\hat{n}_1 - \hat{n}_2\right)\left(\hat{B}_1 - \hat{B}_2\right)+ \frac{1}{2} \left(\hat{n}_1 + \hat{n}_2\right)\left(\hat{B}_1 + \hat{B}_2\right) \\ 
& =  \left(\hat{n}_1 - \hat{n}_2\right) \sum_k \lambda_k \left(\hat{b}_k + \hat{b}_k^\dagger\right) + \left(\hat{n}_1 + \hat{n}_2\right) \sum_k \lambda_k \left(\hat{a}_k + \hat{a}_k^\dagger\right),
\end{align}
where we defined $\lambda_k = \lambda_k'/\sqrt{2}$, $\hat{b}_k = (\hat{b}_{k,1} - \hat{b}_{k,2})/\sqrt{2}$ and $\hat{a}_k = (\hat{b}_{k,1} + \hat{b}_{k,2})/\sqrt{2}$. In terms of the new mode operators, the environment Hamiltonian reads as $\hat{\mathcal{H}}'_E = \sum_k\Omega_k(\hat{b}_k^\dagger\hat{b}_k + \hat{a}^\dagger_k\hat{a}_k)$. We now make use of the single-electron constraint by setting $\hat{n}_1 + \hat{n}_2 = 1$. The total Hamiltonian is thus given, up to a constant, by
\begin{equation}
    \hat{\mathcal{H}}' = \hat{\mathcal{H}} + \sum_k \Omega_k \tilde{a}^\dagger_k \tilde{a}_k,
\end{equation}
where $\tilde{a}_k = \hat{a}_k + \lambda_k$ and $\hat{H}$ is given by Eq.~\eqref{H_set_up}. Since $[\tilde{a}_k,\hat{b}_{k'}^\dagger] = 0$, the second term above describes an independent collection of harmonic oscillators, decoupled completely from the system, which can thus be ignored. We therefore recover the model discussed in the previous section, even though each quantum dot interacts with an independent bath. 

\section{Obtaining steady state results for DQD from quantum master equation}\label{Sec:DQD_RQME}
\subsubsection{The general Redfield Quantum Master Equation and quantum noise}
To obtain the NESS results for DQD-unit without the cavity-unit, we will take the approach of the Redfield Quantum Master Equation (RQME). For an arbitrary set-up of a system connected to a bath, the full set-up can be taken as isolated and described via the full system+bath Hamiltonian
\begin{align}
\label{H_general}
\hat{\mathcal{H}}=\hat{\mathcal{H}}_S+\hat{\mathcal{H}}_{SE}+\hat{\mathcal{H}}_E, ~~\hat{\mathcal{H}}_{SE} = \epsilon \hat{S}\hat{B}
\end{align}
 Here $\hat{\mathcal{H}}_S$ is the system Hamiltonian, $\hat{\mathcal{H}}_E$ and $\hat{\mathcal{H}}_{SE}$ is the system-bath coupling Hamiltonian, $\hat{S}$ is some system operator and $\hat{B}$ is some bath operator and $\epsilon$ is a small parameter controlling the strength of system-bath coupling. Note that the above form of system-bath coupling assume $\hat{S}$ and $\hat{B}$ are Hermitian, which is true in our case. The initial density matrix of the set-up $\rho_{tot}(0)$ is considered to be in product form $\rho_{tot}(0) = \rho(0) \otimes \rho_E$, where $\rho(0)$ is some initial state of the system, and $\rho_E$ is the initial state of the bath which is taken to be the thermal state with respect to the bath Hamiltonian. The Redfield Quantum Master Equation (RQME) is obtained by writing down the equation of motion for the reduced density matrix of the system up to $O(\epsilon^2)$ under Born-Markov approx. If $\langle\hat{B}_\ell\rangle_E=0$, where $\langle...\rangle_E$ refers to the average taken only with respect to bath, is satisfied initially, then, the RQME is given by
\begin{align}
\label{RQME_general}
\frac{\partial\rho}{\partial t} &=i[\rho, \hat{\mathcal{H}}_S] -\epsilon^2 \int_0^\infty dt^{\prime} \Big\{ [\hat{S},\hat{S}(-t^{\prime})\rho(t)]\langle \hat{B}(t^\prime)\hat{B}(0)\rangle_E  +[\rho(t)\hat{S}(-t^{\prime}),\hat{S}]\langle \hat{B}(0)\hat{B}(t^\prime)\rangle_E \Big\}~,  
\end{align}
where $\hat{S}_m(t)=e^{i\hat{\mathcal{H}}_St} \hat{S}_m e^{-i\hat{\mathcal{H}}_S t}$, $\hat{B}_m(t)=e^{i\hat{\mathcal{H}}_Et} \hat{B}_m e^{-i\hat{\mathcal{H}}_E t}$. This gives the leading order dissipative term. 
The evolution equation for expectation value of any system operator $\hat{O}$ is given by
\begin{align}
\label{gen_operator_eqn}
\frac{d\langle \hat{O} \rangle}{dt} =& -i\big\langle [\hat{O},\hat{\mathcal{H}}_S]\big\rangle - \epsilon^2\int_0^\infty dt  \big\langle [ \hat{O}, \hat{S} ]\hat{S}^I(-t) \big\rangle \langle\hat{B}(t)\hat{B}(0)\rangle_E +\int_0^\infty dt  \big\langle  \hat{S}^I(-t) [ \hat{O}, \hat{S} ] \big\rangle \langle\hat{B}(0)\hat{B}(t)\rangle_E,
\end{align}
where $\langle ... \rangle=Tr(\rho ...)$.  Note that if system-bath coupling Hamiltonian is $O(\epsilon)$, the dissipative part of RQME is $O(\epsilon^2)$. This means, following Ref.~\cite{accuracy_of_qme}, the Redfield equation gives results which are correct to   $O(\epsilon^2)$ for the off-diagonal elements of $\rho$ in the eigenbasis of the $\hat{\mathcal{H}}_S$, while it gives results correct to $O(\epsilon^0)$ for the off-diagonal elements of the $\rho$ in the eigenbasis of the $\hat{\mathcal{H}}_S$. In the following, and in the main text, we absorb $\epsilon$ into the definition of the system-bath coupling, which amounts to rescaling the spectral density as $\mathfrak{J}_{ph}(\omega) \to \epsilon^2 \mathfrak{J}_{ph}(\omega)$.

Note that, assuming $\hat{B}$ is Hermitian, the dissipative part in Eqs.~\eqref{RQME_general} and \eqref{gen_operator_eqn} depend on the bath correlation functions $\langle\hat{B}(t)\hat{B}(0)\rangle_E$ and its Hermitian conjugate. This correlation function has the meaning of an effective noise that the system experiences due to the bath. Noise is characterized by its power spectral density, defined as the Fourier transform of $\langle\hat{B}(t)\hat{B}(0)\rangle_E$,
\begin{align}
\label{noise_psd}
& W(\omega) = \int_{-\infty}^{\infty} \frac{dt}{2\pi} e^{i\omega t}\langle\hat{B}(t)\hat{B}(0)\rangle_E,\nonumber \\
&\Rightarrow 
\langle\hat{B}(t)\hat{B}(0)\rangle_E = \int_{-\infty}^{\infty} d\omega W(\omega)e^{-i\omega t} = \int_0^{\infty} 
d\omega \left[W_s(\omega) \cos(\omega t) - i W_a(\omega) \sin(\omega t)\right], \\
& W_s(\omega) = \frac{W(\omega)+W(-\omega)}{2}, ~~ W_a(\omega) = \frac{W(\omega)-W(-\omega)}{2} \nonumber
\end{align}
Here, in the second line, we have broken the inverse Fourier transform into symmetric ($W_s(\omega)$) and anti-symmetric parts ($W_a(\omega)$) with respect to $\omega$. If the anti-symmetric part $W_a(\omega)=0$, then, we see from above equation that $\langle\hat{B}(t)\hat{B}(0)\rangle_E$ is real. This would mean that $[\hat{B}(t),\hat{B}(0)]=0$. So, as far as the noise of the system is concerned, $\hat{B}(t)$ will no longer behave as an operator, but as a classical variable. If, on the other hand, $[\hat{B}(t),\hat{B}(0)]\neq 0$, then $\langle\hat{B}(t)\hat{B}(0)\rangle_E$ will have a complex part. Hence, for such a case, $W_a(\omega)\neq 0$. Thus, existence of the anti-symmetric part of the power-spectral density is a necessary signature of the quantum nature of the noise coming from the environment~\cite{quantum_noise}. In the following, we will see that it is the quantum nature of the noise coming from the bath in our set-up that leads to steady-state coherence.

\subsubsection{Steady-state-coherence in the charge qubit due to quantum noise}
We now simply use Eqs.~\eqref{gen_operator_eqn},~\eqref{RQME_general} for our set-up of interest, the charge qubit given in Eq.~\eqref{charge_qubit}. Comparing Eq.~\eqref{H_general} and Eq.~\eqref{charge_qubit}, we see that
\begin{align}
\hat{S}=\Big[ f_1 \hat{\sigma}_z +f_2 \hat{\sigma}_x\Big],~~\hat{B}=\sum_{s=1}^\infty \lambda_k \left(\hat{b}_{k}^{\dagger} + \hat{b}_{k}\right).
\end{align}
We define the spectral function of the phonon bath
\begin{align}
\label{bath_spectral}
\mathfrak{J}_{ph}(\omega) = \sum_{k=1}^\infty {\lambda_{k}}^2 \delta(\omega - \Omega_{k}).
\end{align}
The effective noise correlation from the bath can be written as
\begin{align}
\label{noise_correlation}
\langle\hat{B}(t)\hat{B}(0)\rangle_E = \int_0^\infty d\omega \mathfrak{J}_{ph}(\omega) \left[ \coth\left(\frac{\beta\omega}{2}\right)
\cos(\omega t)-i\sin(\omega t)\right]
\end{align}
Comparing above equation with Eq.~\eqref{noise_psd}, we see that
\begin{align}
\label{noise_sys_antisym}
W_s(\omega,\beta) = \mathfrak{J}_{ph}(\omega) \coth\left(\frac{\beta\omega}{2}\right),~~W_a(\omega) = \mathfrak{J}_{ph}(\omega), 
\end{align}
where we have added to argument $\beta$ to the definition of symmetric part of $W(\omega)$ to make the temperature dependence explicit.
So, the phononic bath acts as an effective a quantum noise on the system. This is essentially the quantum part of the charge noise. Now, we define the following function
\begin{align}
F_E(\omega_q) = \int_0^\infty dt e^{-i\omega_q t} \langle\hat{B}(t)\hat{B}(0)\rangle_E,
\end{align}
where $\omega_q$ is the qubit frequency in Eq.~\eqref{charge_qubit}. Using Eqs.~\eqref{noise_correlation} and \eqref{noise_sys_antisym}, $F_E(\omega_q)$ can be evaluated as
\begin{align}
\label{def_Fb_Delta}
& F_E(\omega_q) = \frac{\pi}{2}\left[W_s(\omega_q,\beta)-W_a(\omega_q)\right] - i \frac{1}{2}\left[\Delta_s(\omega_q,\beta) + \Delta_a(\omega_q)\right] \nonumber \\ 
& \Delta_s(\omega_q,\beta) = \mathcal{P} \int_0^\infty d\omega W_s(\omega,\beta)\left[\frac{1}{\omega+\omega_q} - \frac{1}{\omega-\omega_q}\right], ~~ \Delta_a(\omega_q) = \mathcal{P} \int_0^\infty d\omega W_a(\omega)\left[\frac{1}{\omega+\omega_q} + \frac{1}{\omega-\omega_q}\right].
\end{align}
In terms of the functions defined above, the evolution expectation values of $\langle\hat{\sigma}_x\rangle$, $\langle\hat{\sigma}_y\rangle$, $\langle\hat{\sigma}_z\rangle$ can be written down using Eq.~\eqref{gen_operator_eqn},
\begin{align}
& \frac{d\langle\hat{\sigma}_x\rangle }{dt}=-\omega_q \langle\hat{\sigma}_y\rangle + 2\pi f_1 f_2 \left(W_s(\omega_q,\beta)\langle\hat{\sigma}_z\rangle+W_a(\omega_q)\right) \nonumber \\
& \frac{d\langle\hat{\sigma}_y\rangle }{dt}= \left(\omega_q + 2 f_2^2 \Delta_s(\omega_q,\beta)\right) \langle\hat{\sigma}_x\rangle - 2\pi f_2^2 W_s(\omega_q,\beta)\langle\hat{\sigma}_y\rangle - 2 f_1 f_2 \left[-\Delta_s (\omega_q,\beta) \langle\hat{\sigma}_z\rangle+\Delta_a(\omega_q)-\Delta_a(0)\right] \nonumber \\
& \frac{d\langle\hat{\sigma}_z\rangle }{dt}=- 2\pi  f_2^2 \left(W_s(\omega_q,\beta)\langle\hat{\sigma}_z\rangle+W_a(\omega_q)\right).
\end{align}
The steady state solution is obtained by setting the LHS of the above equations to zero. Thus, we obtain the steady state solutions up to leading order in system-bath coupling,
\begin{align}
\label{steady_state_soln}
&\langle\hat{\sigma}_z\rangle_0 = -\frac{W_a(\omega_q)}{W_s(\omega_q,\beta)}=-\tanh\left(\frac{\beta\omega_q}{2}\right) , ~~\langle\hat{\sigma}_y\rangle=0,~~ \langle\hat{\sigma}_x\rangle = \frac{2f_1 f_2}{\omega_q}\left[\Delta_s(\omega_q,\beta) \langle\hat{\sigma}_z\rangle + \Delta_a(\omega_q)-\Delta_a(0)\right],
\end{align}
where we have used $\langle\hat{\sigma}_z\rangle_0$ to highlight that this is the zeroth order result in system-bath coupling, while $\langle\hat{\sigma}_x\rangle$ and $\langle\hat{\sigma}_y\rangle$ are second order in system-bath coupling.
The above equations show very interesting physics. If the charge noise was classical, $W_a(\omega)=\Delta_a(\omega)=0$. In that case, we would have $\langle\hat{\sigma}_z\rangle_0=\langle\hat{\sigma}_y\rangle=\langle\hat{\sigma}_x\rangle=0$ in the steady state. So a classical noise would take any given initial state of the system to a completely mixed state. If, on the other hand, the noise from the environment is quantum, as in our case, $\langle\hat{\sigma}_x\rangle\neq 0$ in the steady state. So, no matter what state the system was in initially, in the steady state, coherence will be generated due to the quantum noise coming from the environment.

It is important to note that the fact that $\langle\hat{\sigma}_y\rangle=0$ in the equilibrium steady state is physically consistent. This is because, 
\begin{align}
\hat{\sigma}_y = -i(\hat{A}_1^\dagger\hat{A}_2-\hat{A}_2^\dagger\hat{A}_1)= i(\hat{c}_1^\dagger\hat{c}_2 + \hat{c}_2^\dagger\hat{c}_1),
\end{align} 
it proportional to the particle current operator. Since the system is in equilibrium, and the full system+bath Hamiltonian has time-reversal symmetry, there can be no particle current in the steady state. Thus, $\langle\hat{\sigma}_y\rangle=0$, as we have found above, is mandatory for respecting second law of thermodynamics. 

As mentioned before, the second order perturbative quantum master gives the the diagonal elements of the density matrix correct to zeroth order. So, in above equation we get  $\langle\hat{\sigma}_z\rangle _0= -\tanh\left(\frac{\beta\omega_q}{2}\right)$, which is the zeroth order result. At low temperatures, this essentially violates the necessary condition for a qubit that $r=\sqrt{\langle\hat{\sigma}_x\rangle^2 + \langle\hat{\sigma}_y\rangle^2 + \langle\hat{\sigma}_z\rangle^2}\leq 1$. So, for the results to be meaningful, we need to calculate the second order correction to $\langle\hat{\sigma}_z\rangle$. This can be done by deriving the quantum master equation up to higher order. However, that is quite cumbersome. So, instead, we use the fact that the steady state of the system in equilibrium should be given by the marginal of the global thermal state, i.e., $\rho=Tr_E(exp(-\beta \hat{\mathcal{H}})/Z)$, where $\hat{\mathcal{H}}$ is the total Hamiltonian of the set-up (see Eq.~\eqref{H_set_up}). In the next section, we calculate $\langle\hat{\sigma}_x\rangle $ and $\langle\hat{\sigma}_z\rangle $ by perturbative expansion of global thermal state. We show that $\langle\hat{\sigma}_x\rangle $ obtained is the same as Eq.~\eqref{steady_state_soln}, while we can obtain the next order correction to $\langle\hat{\sigma}_z\rangle $.
 
\section{DQD results from perturbation expansion of a global thermal state}
\subsubsection{perturbation expansion of a global thermal state}\label{Sec:DQD_global_thermal_state}
Let us write the total Hamiltonian as $\hat{\mathcal{H}} = \mathcal{\hat{\mathcal{H}}}_S + \hat{\mathcal{H}}_E + \mathcal{\hat{\mathcal{H}}}_{SE}$, where
\begin{align}\label{H_sys}
\mathcal{\hat{\mathcal{H}}}_S & = \frac{\omega_q}{2} \hat{\sigma}_z,\\
\label{H_bath}
\hat{\mathcal{H}}_E & = \sum_{k}\Omega_k \bdag_k \b_k, \\
\label{H_int}
\mathcal{\hat{\mathcal{H}}}_{SE} & = \left ( f_1 \hat{\sigma}_z + f_2 \hat{\sigma}_x\right )\sum_k \lambda_k \left ( \b_k + \bdag_k \right ).
\end{align}
Our goal is to compute the reduced density matrix
\begin{equation}\label{marginal_def}
\rho_S = \Tr_E\left [\frac{\ee^{-\beta \hat{\mathcal{H}}}}{Z}\right ],
\end{equation}
up to second order in $\mathcal{\hat{\mathcal{H}}}_{SE}$, where $Z = \Tr [\ee^{-\beta \hat{\mathcal{H}}} ]$. For this purpose, let us define $\hat{X}(\beta) =  \ee^{\beta \hat{\mathcal{H}}_0} \ee^{-\beta \hat{\mathcal{H}}}$, where $\hat{\mathcal{H}}_0 = \mathcal{\hat{\mathcal{H}}}_S + \hat{\mathcal{H}}_E$. This is the solution of the differential equation
\begin{equation}\label{diff_eqn}
\frac{\partial \hat{X}}{\partial \beta} = - \mathcal{\tilde{H}}_{SE}(\beta) \hat{X}(\beta), 
\end{equation}
where $\mathcal{\tilde{H}}_{SE}(\beta) = \ee^{\beta \hat{\mathcal{H}}_0}\mathcal{\hat{\mathcal{H}}}_{SE} \ee^{-\beta \hat{\mathcal{H}}_0}$. Using the boundary condition $\hat{X}(0) = \mathbbm{1}$, this equation has a well-known solution in terms of a time-ordered exponential:
\begin{equation}\label{time_exp_solution}
\hat{X}(\beta) = {\rm T} \exp \left [-\int_0^\beta \dd \tau\, \mathcal{\tilde{H}}_{SE}(\tau)\right ].
\end{equation}
Truncating this at second order in $\mathcal{\tilde{H}}_{SE}$, we find 
\begin{equation}\label{X_second_order}
\hat{X}(\beta)  \approx  \mathbbm{1} -  \int_0^\beta\dd \tau\, \mathcal{\tilde{H}}_{SE}(\tau)  + \int_0^\beta\dd \tau\int_0^\tau\dd \tau'\, \mathcal{\tilde{H}}_{SE}(\tau) \mathcal{\tilde{H}}_{SE}(\tau').
\end{equation}
We now recover the equilibrium density matrix from the identity
\begin{equation}\label{equ_density}
\ee^{-\beta \hat{\mathcal{H}}} = \frac{1}{2}\left [ \ee^{-\beta\hat{\mathcal{H}}_0} \hat{X}(\beta) + \hat{X}^\dagger(\beta)\ee^{-\beta \hat{\mathcal{H}}_0}  \right ] ,
\end{equation}
which enforces hermiticity of the density matrix even with the approximated expression~\eqref{X_second_order}.

Now, let us define the bare expectation value $\langle \bullet \rangle_0 = \Tr[ \bullet \ee^{-\beta \hat{\mathcal{H}}_0}/Z_0]$, where $Z_0 = \Tr[ \ee^{-\beta \hat{\mathcal{H}}_0}]$ is the bare partition function, which factorizes into system and bath contributions $Z_{S,B} = \Tr[\ee^{-\beta \hat{\mathcal{H}}_{S,B}}]$.. We also write $\mathcal{\hat{\mathcal{H}}}_{SE} = \hat{S} \hat{B}$, where $\hat{S} =  f_1 \sg_z+ f_2 \sg_x$ and $\hat{B}=\sum_k g_k  ( \b_k + \bdag_k  )$. Since $\Tr_E[\mathcal{\hat{\mathcal{H}}}_{SE}\ee^{-\beta\hat{\mathcal{H}}_0}]=0$, we obtain the approximate partition function
\begin{equation}\label{partition_function}
Z \approx Z_0\left [1  + \int_0^\beta \dd \tau \int_0^\tau \dd\tau'\, \Re \left \langle \mathcal{\tilde{H}}_{SE}(\tau) \mathcal{\tilde{H}}_{SE}(\tau') \right \rangle_0\right ].
\end{equation}
Similarly, we obtain the numerator of Eq.~\eqref{marginal_def} as
\begin{equation}\label{numerator}
\Tr_E \left [ \ee^{-\beta\hat{\mathcal{H}}} \right ] \approx Z_E \ee^{-\beta \mathcal{\hat{\mathcal{H}}}_S} + \frac{Z_E}{2} \int_0^\beta \dd \tau \int_0^\tau \dd\tau'\, \left [ \ee^{-\beta \mathcal{\hat{\mathcal{H}}}_S} \tilde{S}(\tau) \tilde{S}(\tau')\left \langle \tilde{B}(\tau)\tilde{B}(\tau')\right \rangle_0 + \rm h.c.\right ],
\end{equation}
where $\tilde{S}(\tau) = \ee^{\tau \mathcal{\hat{\mathcal{H}}}_S}\hat{S} \ee^{-\tau \mathcal{\hat{\mathcal{H}}}_S}$, $\tilde{B}(\tau) = \ee^{\tau \hat{\mathcal{H}}_E}\hat{B} \ee^{-\tau \hat{\mathcal{H}}_E}$.
Dividing one by the other, we get the following result at second order:
\begin{equation}\label{marginal_2ndorder}
\rho_S = \frac{\ee^{-\beta \mathcal{\hat{\mathcal{H}}}_S}}{Z_S} \left [1 -   \int_0^\beta \dd \tau \int_0^\tau \dd\tau'\, \phi(\tau-\tau')\Re \left \langle \tilde{S}(\tau) \tilde{S}(\tau') \right \rangle_0\right ] + \frac{1}{2}\int_0^\beta \dd \tau \int_0^\tau \dd\tau'\,  \phi(\tau-\tau')\left [ \frac{\ee^{-\beta \mathcal{\hat{\mathcal{H}}}_S}}{Z_S} \tilde{S}(\tau) \tilde{S}(\tau') + \rm h.c.\right ].
\end{equation}
Here we defined the bath correlation function $  \phi(\tau-\tau')=\langle\tilde{B}(\tau) \tilde{B}(\tau') \rangle $. 

\subsubsection{Results for the DQD}
In our set-up $\phi(\tau-\tau')$ is given explicitly by
\begin{align}\label{bath_correlation_function}
\phi(\tau)  = \int_0^\infty\dw{\omega} \mathfrak{J}(\omega) \left [ \coth(\beta\omega/2) \cosh(\omega\tau) - \sinh(\omega\tau) \right ].
\end{align}
Furthermore, the relevant product of system operators is given by
\begin{align}\label{sys_op_product}
\tilde{S}(\tau) \tilde{S}(\tau')  & = f_1^2 + f_2^2 \cosh[\omega_q(\tau-\tau')] + f_2^2 \sinh[\omega_q(\tau-\tau')]\sg_z \notag \\ 
& \quad + f_1 f_2 [\sinh(\omega_q\tau') - \sinh(\omega_q \tau)]\sg_x + \ii f_1 f_2[\cosh(\omega_q\tau')- \cosh(\omega_q\tau)]\sg_y.
\end{align}
We thus find that
\begin{equation}\label{sys_corr}
 \left \langle \tilde{S}(\tau) \tilde{S}(\tau') \right \rangle_0 = f_1^2 + f_2^2\cosh[\omega_q(\tau-\tau')] - f_2^2\tanh(\beta\omega_q/2) \sinh[\omega_q(\tau-\tau')].
\end{equation}

Using the above results, it is straightforward to deduce that $\langle \sg_y\rangle = 0$. Note that the hermitian conjugate term in Eq.~\eqref{marginal_2ndorder} is essential to obtain this result. For the relevant steady-state coherence, we find
\begin{align}\label{SSC_result}
\langle \sg_x\rangle  & = f_1 f_2 \int_0^\beta \dd\tau \int_0^\tau\dd\tau'\phi(\tau-\tau') \left \lbrace \tanh(\beta\omega_q/2) [\cosh(\omega_q\tau) - \cosh(\omega_q\tau')] - [\sinh(\omega_q\tau) - \sinh(\omega_q\tau')] \right \rbrace, \notag \\
\end{align}
It can be checked that after plugging in Eq.~\eqref{bath_correlation_function} and carrying out the imaginary-time integrals explicitly, we obtain the same result as in Eq.~\eqref{steady_state_soln}. We also obtain a correction to the bare population inversion $\langle \sg_z\rangle_0 = -\tanh(\beta\omega_q/2)$ given by
\begin{align}\label{sigZ_result}
\langle \sg_z\rangle - \langle \sg_z\rangle_0 & = f_2^2{\rm sech}^2(\beta\omega_q/2)\int_0^\beta\dd\tau\int_0^\tau\dd\tau'\, \phi(\tau-\tau') \sinh[\omega_q(\tau-\tau')],\notag \\
& = 2 f_2^2\int\dw{\omega} \mathfrak{J}_{ph}(\omega) \frac{\coth(\beta\omega/2)\left[2(\omega^2+\omega_q^2)\tanh(\beta\omega_q/2) + \beta\omega_q(\omega^2-\omega_q^2){\rm sech}^2 (\beta\omega_q/2)\right] - 4\omega\omega_q}{(\omega^2-\omega_q^2)^2}.
\end{align}
Defining the following functions,
\begin{align}
\Delta_{\pm}(\omega_q,\beta) =\mathcal{P} \int_0^\infty \frac{d\omega \mathfrak{J}_{ph}(\omega)}{(\omega\pm \omega_q)^2} \left[\tanh(\frac{\omega_q\beta}{2}) \coth(\frac{\omega\beta}{2})\pm 1\right],
\end{align}
the final result in Eq.~\eqref{sigZ_result} can be simplified to
\begin{align}
\langle\hat{\sigma}_z\rangle =- \tanh\left(\frac{\beta \omega_q}{2}\right)  + f_2^2\left[\Delta_-(\omega_q,\beta)+\Delta_+(\omega_q,\beta)-\frac{\beta}{2}\textrm{sech}^2\left(\frac{\beta \omega_q}{2}\right) \Delta_s(\omega_q,\beta)\right].
\end{align}
Here $\Delta_s(\omega_q,\beta)$ is as defined in Eq.~\eqref{def_Fb_Delta}.

\section{Distinguishing from effects of renormalized qubit parameters} \label{Sec:not_renormalized_qubit}
In above, we have shown that the equilibrium steady state of the DQD charge qubit will have coherence in the eigenbasis of the system Hamiltonian due to quantum noise from the phonons of the substrate. It is well-known that when connected to a bath the parameters of the system Hamiltonian gets renormalized. The eigenbasis of the renormalized Hamiltonian may not be the same as that of the original Hamiltonian. The natural question to ask is whether the state of the qubit with coherence is actually a thermal state of the renormalized Hamiltonian. In the following, we show that this is \emph{not true}.

One standard way to find the renormalized Hamiltonian is to obtain the so-called `Lamb-shift' Hamiltonian from the quantum master equation Eq.~\eqref{RQME_general}. In order to do this, we have to convert Eq.~\eqref{RQME_general} to the standard Gorini-Kossakowski-Sudarshan-Lindblad (GKSL) form. For a quantum master equation describing a qubit, the general GKSL form of the quantum master equation can be written as
\begin{align}
\frac{\partial\rho}{\partial t} &=i[\rho, \hat{\mathcal{H}}_{eff} ]+\mathcal{L}(\rho),~~\mathcal{L}(\rho)=\sum_{p,q=x,y,z} M_{pq}\left(\hat{\sigma}_p \rho \hat{\sigma}_q - \frac{1}{2}\Big\{\rho,\hat{\sigma}_q \hat{\sigma}_p\Big\}\right),~~ \hat{\mathcal{H}}_{eff}=\hat{\mathcal{H}}_S + \hat{\mathcal{H}}_{LS},
\end{align}
where $\{\hat{f},\hat{g} \}=\hat{f}\hat{g}+\hat{g}\hat{f}$ is the anti-commutator. In the GKSL form, $\mathcal{L}(\rho)$ gives the non-unitary part of the quantum master equation due to the presence of the bath, while $\hat{\mathcal{H}}_{LS}$ gives the `Lamb-shift' Hamiltonian coming due to the presence of the bath. We want to see if the equilibrium steady state with coherence can be interpreted as a thermal state of $\hat{\mathcal{H}}_{eff}$  with inverse temperature $\beta$.
To convert Eq.~\eqref{RQME_general} to the above form, we first write it as
\begin{align}
\frac{\partial\rho}{\partial t} &=i[\rho, \hat{\mathcal{H}}_S] +\epsilon^2 \mathcal{D}(\rho),~~ \mathcal{D}(\rho)= -\Big\{ [\hat{S},\hat{R}\rho(t)]  +[\rho(t)\hat{R}^\dagger,\hat{S}]\Big\},~~ \hat{R} = \int_{0}^{\infty} dt \hat{S}(-t) \langle \hat{B}(t) \hat{B}(0) \rangle_E.
\end{align}
In above, we have used the fact that $\hat{S}$ and $\hat{B}$ are Hermitian. For our system, we have,
\begin{align}
\label{def_R}
\hat{S}=f_1 \hat{\sigma}_x + f_2 \hat{\sigma}_z,~~\hat{R}=f_1 \left[ \left(\frac{F_E(\omega_q)+F_E(-\omega_q)}{2}\right)\hat{\sigma}_x + i\left(\frac{F_E(\omega_q)-F_E(-\omega_q)}{2}\right)\hat{\sigma}_y\right] + f_2 F_E(0)\hat{\sigma}_z 
\end{align}
We can now write $\mathcal{D}(\rho)$ in the following form, 
\begin{align}
\mathcal{D}(\rho) = \mathcal{L}(\rho) + i\left[\rho, \hat{\mathcal{H}}_{LS}  \right],~~\mathcal{L}(\rho)=\hat{R}\rho \hat{S} - \frac{1}{2} \Big\{\rho, \hat{S}\hat{R}\Big\} + \hat{S}\rho \hat{R}^\dagger - \frac{1}{2} \Big\{\rho, \hat{R}^\dagger \hat{S}\Big\},~~\hat{\mathcal{H}}_{LS}=\frac{\hat{R}^\dagger\hat{S}-\hat{S}\hat{R}}{2}. 
\end{align}
 Using Eq.~\eqref{def_R}, $\mathcal{L}(\rho)$ can be expanded to explicitly have the GKSL form. Using the same, the `Lamb-shift' Hamiltonian is given by,
\begin{align}
\hat{\mathcal{H}}_{LS} = f_1 f_2 \Big [ \frac{1}{2}\Delta_s(\omega_q,\beta) \hat{\sigma}_x - \frac{\pi}{2}W_s(\omega_q, \beta) \hat{\sigma}_y \Big ] - \frac{f_1^2 }{2}\Delta_s(\omega_q,\beta) \hat{\sigma}_z.
\end{align}
We immediately see that, for $f_1 f_2 \neq 0$,  $\hat{\mathcal{H}}_{LS}$ has a $\hat{\sigma}_y$ component. \emph{Thus, a thermal state of $\hat{\mathcal{H}}_{eff}$, $\rho_{eff}\propto \exp(-\beta\hat{\mathcal{H}}_{eff})$ will have a non-zero expectation value of $\hat{\sigma}_y$, i.e, $Tr\Big(\rho_{eff}\hat{\sigma}_y\Big)\neq 0$. As discussed before, such a state cannot be the steady state of our system because it would violate the second law of thermodynamics.}

Having shown that the steady state with coherence cannot be described as a thermal state with respect to a `Lamb-shifted' Hamiltonian, let us now ask if it at all possible to find a \emph{meaningful} Hamiltonian with respect to which the steady state could be described as a thermal state. To this end, we note that, any state of a qubit can be written as a thermal state of with respect to a renormalized Hamiltonian with a given temperature. To see this note that the eigenvalues of any general state of a qubit are given by $\frac{1+r}{2}$ and $\frac{1-r}{2}$, where $r=\sqrt{\langle \hat{\sigma}_x\rangle^2+\langle \hat{\sigma}_y\rangle^2+\langle \hat{\sigma}_z\rangle^2}$, is the radius of the Bloch sphere. This state is a `thermal state' with respect to a qubit with the qubit frequency $\tilde{\omega}_q$ given by
\begin{align}
\label{def_tilde_omega_q}
\frac{1+r}{1-r}=e^{\beta\tilde{\omega}_q}\Rightarrow \tilde{\omega}_q=\frac{1}{\beta}\log\left(\frac{1+r}{1-r}\right).
\end{align}
The above way to define the renormalized qubit in a thermal state of given temperature is always possible. However, note that, in doing so, in general, the qubit frequency defined may become temperature dependent, making $\tilde{\hat{\mathcal{H}}}_S$ a function of temperature. This makes such definition completely artifical. Thus, the only meaningful case in this respect is when $r$ is such that $\tilde{\omega}_q$ is independent of $\beta$. Only in that case, the above will point to the fact that the given density matrix is indeed a true thermal state of the qubit with frequency $\tilde{\omega}_q$. For the steady state of our system at $\beta\omega_q\gg 1$,
\begin{align}
\label{sigmaxz_zero_temp}
\langle\hat{\sigma}_x\rangle =-2\sin 2\theta\int_0^\infty  d\omega\frac{\mathfrak{J}_{ph}(\omega)}{\omega(\omega+\omega_q)},~~\langle\hat{\sigma}_z\rangle = -1 + 2\sin^2\theta\int_0^\infty d\omega \frac{\mathfrak{J}_{ph}(\omega)}{(\omega+\omega_q)^2},
\end{align}  
and $\langle\hat{\sigma}_y\rangle=0$. Thus, $r$ becomes independent of $\beta$ for $\beta \omega_q \gg 1$. So, if we were to define an effective Hamiltonian of the qubit, with respect to which, the steady state is a thermal state, its frequency $\tilde{\omega}_q \propto 1/\beta$ (from Eq.~\eqref{def_tilde_omega_q}). So, it is impossible to find any \emph{meaningful} qubit Hamiltonian with respect to which the steady state could be described as a thermal state.

Hence, in this section, we have shown that the qubit state with coherence in the energy eigenbasis cannot be explained as an effect of simple renormalization of qubit parameters due to the presence of phonons.

\section{Quantum noise generates coherence in any qubit}\label{Sec:generality_of_results}
Even though in above we have specifically focused on one experimental platform, the charge qubit, our results are valid for any arbitrary qubit (flux qubits, superconducting qubits, spin qubits etc.). Any arbitrary qubit Hamiltonian, preserving time-reversal symmetry, and in presence of noise can be written as
\begin{align}
\hat{H} = \left(\frac{\varepsilon}{2}+\lambda_{\varepsilon}\hat{B}\right) \hat{\tau}_z +\left( t_c+\lambda_{t_c}\hat{B}\right) \hat{\tau}_x,
\end{align} 
where $\hat{\tau}_z$ and $\hat{\tau}_x$ are Pauli spin matrices, $\hat{B}$ is the `noise' operator embodying, $\lambda_{\varepsilon}$ is the strength of the `noise' in detuning, $\lambda_{t_c}$ is the strength of the `noise' in the off-diagonal term.  For simplicity, we have assumed that the sources of `noise' in both diagonal and off-diagonal terms are the same.  As discussed before, the `noise' operator is defined by their power-spectral density (see Eq.~\eqref{noise_psd}).  For quantum noise, the `noise' can be assumed to arise from a bosonic bath, as before, which essentially amounts to choosing the same power-spectral density.  
The system Hamiltonian, i.e, the Hamiltonian without the `noise' operators can be diagonalized by a rotation in the space of Pauli matrices,
\begin{align}
&\left(
\begin{array}{c}
\hat{\sigma}_x\\
\hat{\sigma}_z\\
\end{array}
\right)= \left(
\begin{array}{cc}
\cos(2\theta) & -\sin(2\theta)\\
\sin(2\theta) & \cos(2\theta)\\
\end{array}
\right) \left(
\begin{array}{c}
\hat{\tau}_x\\
\hat{\tau}_z\\
\end{array}
\right), ~~\theta = \frac{1}{2}\tan^{-1}\left(\frac{2 t_c}{\varepsilon}\right).
\end{align}
After going to the eigenbasis of the system Hamiltonian, we now have,
\begin{align}
\hat{H}=\frac{\omega_q}{2}\hat{\sigma}_z + (\tilde{f}_1 \hat{\sigma}_z +\tilde{f}_2 \hat{\sigma}_x)\hat{B},~~ \tilde{f}_1=\lambda_\varepsilon \cos(2\theta) + \lambda_{t_c} \sin(2\theta),~~ \tilde{f}_2=\lambda_{t_c} \cos(2\theta) - \lambda_\varepsilon  \sin(2\theta).
\end{align}
This is a Hamiltonian exactly of the same form as Eq.~\eqref{charge_qubit}. Assuming the couplings to`noise'  to be small, the exact same analysis as before holds, and thereby, Eq.~\eqref{steady_state_soln} holds. \emph{ So, if  $\tilde{f}_1\tilde{f}_2 \neq 0$, quantum noise in qubit parameters leads to generation of coherence in the eigenbasis of the qubit Hamiltonian of the steady state of any qubit.} On the other hand, any classical noise destroys coherence. When both are present, classical noise reduces the amount of coherence.

\section{Master equation in the presence of fermionic leads}\label{Sec:DQD_with_leads}

A DQD set-up usually has fermionic leads attached to the fermionic sites. With the fermionic leads, the number of particles in the DQD is not conserved. The DQD can only be considered as a qubit in the single-particle sector. We show here that the DQD will have coherence in energy eigenbasis in the equilibrium steady state, as long as the equilibrium steady state has a single particle, even in the presence of fermionic leads. The amount of coherence is the same as that in absence of the fermionic leads. 

To show this,  we couple each site of the DQD to the its own fermionic lead.   The total Hamiltonian of the set-up is now given by,
\begin{align}
\hat{\mathcal{H}}_{Sf} = \sum_{\ell=1,2}\sum_{r=1}^{\infty} \gamma_{r\ell} \left(\hat{c}_\ell^\dagger\hat{B}_{\ell r} + \hat{B}_{\ell r}^\dagger \hat{c}_\ell\right),~~
\hat{\mathcal{H}}_{f} = \sum_{\ell=1,2}\sum_{r=1}^{\infty}\mathcal{E}_{\ell r} \hat{B}_{\ell r}^\dagger \hat{B}_{\ell r},~~  \hat{\mathcal{H}}=\hat{\mathcal{H}}_{S}+\hat{\mathcal{H}}_{SE}+\hat{\mathcal{H}}_{E}+\hat{\mathcal{H}}_{Sf}+\hat{\mathcal{H}}_{f},
\end{align} 
where the fermionic lead is modelled by infinite number of fermionic modes, and $\hat{B}_{\ell r}$ is the annihilation operator for the $r$th mode of the fermionic lead attached to the $\ell$th DQD site. For simplicity we will assume that the couplings of both fermionic sites to the fermionic leads are the same, and the leads have same spectral functions. We will make the so-called wide-band-limit approximation of a constant spectral function of the lead,
\begin{align}
\mathfrak{J}^f_\ell(\omega) = 2\pi \sum_{r=1}^\infty \gamma_{\ell r}^2 \delta(\omega-\mathcal{E}_{\ell r}), ~~\mathfrak{J}^f_1(\omega)= \mathfrak{J}^f_2(\omega)=
\begin{cases}
\Gamma,~~\forall~~-\Lambda\leq \Gamma \leq \Lambda \\
0,~~\textrm{otherwise}.
\end{cases}
\end{align}
Here $\Lambda$ is the cut-off frequency, which will be taken to be large.
Since we are concerned with the equilibrium set-up, each fermionic lead is assumed to be at the same temperature $\beta$ as the phononic bath, and have a chemical potential $\mu$.
When couplings to the fermionic leads are also weak, RQME can be derived for the set-up. In order to write down the RQME, we first define the following notations
\begin{align}
&m=
\left(
\begin{array}{cc}
f_1 & f_2\\
f_2 & -f_1\\
\end{array}
\right),~~  \mathfrak{n}^F(\omega) = \frac{1}{e^{\beta(\omega-\mu)}+1}\\
&F(\omega) =  \frac{\Gamma \mathfrak{n}^F(\omega)}{2},~~F^{\Delta}(\omega) = \mathcal{P}\int_{-\Lambda}^\Lambda \frac{d\omega^\prime}{2\pi} \frac{\Gamma \mathfrak{n}^F(\omega^\prime)}{\omega^\prime -\omega}, ~~ f^{\Delta}(\omega) = \mathcal{P}\int_{-\Lambda}^\Lambda \frac{d\omega^\prime}{2\pi} \frac{\Gamma}{\omega^\prime -\omega}.
\end{align}
The RQME is given by,
\begin{align}
\label{RQME_with_leads}
&\frac{\partial \rho}{\partial t} = i[\rho, \hat{\mathcal{H}}_S] -\epsilon^2  \mathcal{L}_{ph} \rho  -\epsilon^2  \mathcal{L}_{f} \rho,  \\
& \mathcal{L}_{ph} \rho = \sum_{\alpha,\nu,\gamma,\delta} \left( m_{\alpha \nu} m_{\gamma \delta} [\hat{A}_\gamma^\dagger \hat{A}_\delta, \hat{A}_\alpha^\dagger \hat{A}_\nu \rho] F_E(\omega_\alpha^f-\omega_\nu^f) + h.c \right), \nonumber \\
& \mathcal{L}_{f} \rho = \sum_{\nu} \left( [\hat{A}_\nu^\dagger, \hat{G}_{\nu} \hat{A}_\nu \rho] + [\rho \hat{F}_{\nu}\hat{A}_\nu, \hat{A}_\nu^\dagger] + h.c \right) \nonumber,
\end{align} 
where $h.c.$ refers to Hermitian conjugate, and 
\begin{align}
&\omega_1^f=\frac{\omega_q}{2}, ~~\omega_2^f=-\frac{\omega_q}{2}, ~~\hat{G}_{\nu} = \Gamma+i f^{\Delta}(\omega_\nu) -\hat{F}_{\nu}, \nonumber \\
& \hat{F}_{\nu} = \left(F(\omega_\nu)+i F^\Delta(\omega_\nu)\right)(1-\hat{N}_{\bar{\nu}}) + \hat{N}_{\bar{\nu}} \left(F(\omega_\nu + V)+iF^\Delta(\omega_\nu + V)\right), 
\end{align}
Here $\bar{\nu}$ is the complement of $\nu$, i.e, if $\nu=1$, $\bar{\nu}=2$ and vice-versa. Note that in Eq.~\eqref{RQME_with_leads} we have kept the perturbative expansion parameter $\epsilon$ explicit. In the main text, this small parameter is absorbed into the system bath coupling by the rescaling $\mathfrak{J}_{ph}(\omega)\to \epsilon^2\mathfrak{J}_{ph}(\omega)$, $\mathfrak{J}^{f}_\ell(\omega)\to\epsilon^2 \mathfrak{J}^{f}_\ell(\omega)$.

\section{Technical aside: validity of results from second order quantum master equation}\label{Sec:Validity_of_RQME}

In above, we have used quantum master equations derived by second order perturbation theory in system-bath coupling. The validity of results from such an equation has already been commented upon. Here we discuss that in more detail, explicitly checking the validity of the RQME with fermionic leads, Eq.~\eqref{RQME_with_leads}.

The RQME is an equation of the form
\begin{align}
\frac{\partial \rho}{\partial t} =L\rho,~~ L\rho= (L_0+ \epsilon^2 L_2)\rho,~~L_0\rho=i[\rho,\hat{\mathcal{H}}_S],~~ L_2\rho=-\mathcal{L}_{ph} \rho -\mathcal{L}_{f} \rho.
\end{align}
In other words, the superoperator $L$ is obtained perturbatively up to second order in system-bath coupling. However, the solution of the above equation is 
\begin{align}
\rho(t) = e^{L t} \rho(0),
\end{align}
which contains all orders in $\epsilon$. Clearly, the obtained result cannot be correct to all orders in $\epsilon$. To what order in $\epsilon$ can the result be trusted? This question was succinctly answered in Ref.~\cite{accuracy_of_qme}. From their result, the diagonal elements of $\rho$ in the eigenbasis of the system Hamiltonian are given correctly up to $O(\epsilon^0)$, whereas, the off-diagonal elements are given correctly up to $O(\epsilon^2)$. Intuitively, this result says only the leading order term arising due to system-bath coupling is accurate. This very important result, though rather under-appreciated in literature, is the reason for the well-known positivity violation of the RQME. This typically happens when one or more of leading order contributions become zero. We have already seen this in Eq.~\eqref{steady_state_soln} in absence of fermionic leads. At low temperatures, the RQME gives $\langle\hat{\sigma}_z\rangle=-1$, and also non-zero value of $\langle \hat{\sigma}_x \rangle$. Taken at face value, this would make $r>1$, a physically impossible situation pointing to positivity violation. However, once we understand the origin of this discrepancy, it becomes clear that this is an artefact of obtaining the QME up to second order, and higher order terms will correct it. This happens because the higher energy state is empty, making the leading order contribution zero. This is why, while using an approximate QME to describe experimental set-ups, one must be careful about the validity of the answers obtained from it. In this work, we have obtained the higher order term in $\langle\hat{\sigma}_z\rangle$ via perturbative expansion of the global thermal state instead. The equivalence between results the perturbative expansion of the global thermal state and the result from perturbative quantum master equation is discussed in Refs.~\cite{Juzar2012,Juzar2013,Juzar2017}.

\begin{figure}
\includegraphics[scale=0.5]{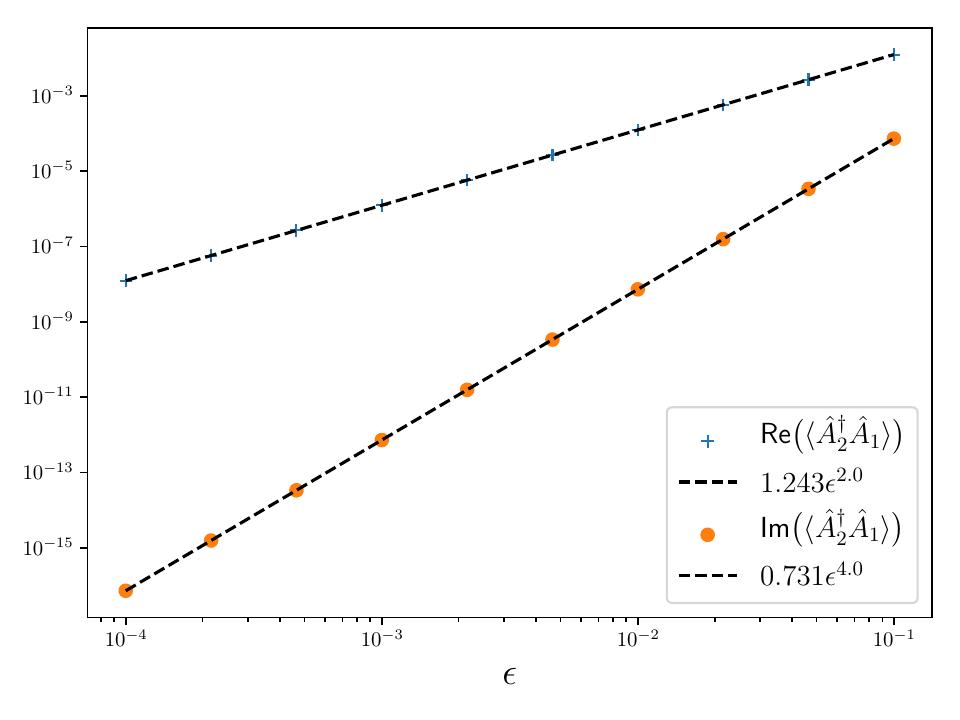} 
\caption{(color online) Scaling of real and imaginary parts of the equilibrium steady state value of $\langle\hat{A}_1^\dagger\hat{A}_2\rangle$ as obtained from Eq.~\eqref{RQME_with_leads} with the small parameter $\epsilon$.  Parameters for both plots: $\varepsilon=1$, $t_c=0.5$, $\mu=0$, $V=5$, $\beta=10$, $\gamma_b=3$, $\omega_{max}=10$, $\Gamma=6$, $\Lambda=400$.  All energies are measured in units of $\omega_c$, which is set to $1$.}
\label{fig:RQME_check} 
\end{figure}

Yet another important application of the results of Ref.~\cite{accuracy_of_qme} comes in the calculation in presence of fermionic leads. A direct calculation of the equilibrium steady state value of $\langle\hat{A}_1^\dagger\hat{A}_2\rangle$ from Eq.~\eqref{RQME_with_leads} gives a real and an imaginary part. This, once again, is a physically inconsistent situation since imaginary part of $\langle\hat{A}_1^\dagger\hat{A}_2\rangle$ is proportional to particle current which should be zero in equilibrium steady state. The resolution of this comes once we look at the scaling of real and imaginary parts of  $\langle\hat{A}_1^\dagger\hat{A}_2\rangle$ with $\epsilon$. This is shown in Fig.~\ref{fig:RQME_check}. We see clearly that 
\begin{align}
\textrm{Re}\Big(\langle\hat{A}_1^\dagger\hat{A}_2\rangle\Big) \propto \epsilon^2, ~~\textrm{Im}\Big(\langle\hat{A}_1^\dagger\hat{A}_2\rangle\Big) \propto \epsilon^4.
\end{align}
Thus, the imaginary part of the coherence in the eigenbasis of the system Hamiltonian scales as $\epsilon^4$ instead of $\epsilon^2$. By previous discussion, this result shows that the non-zero value of the imaginary part is an artefact of the RQME coming from the fact that the $O(\epsilon^2)$ term is zero. On the other hand, the real part of coherence in the eigenbasis of the system scales as $\epsilon^2$ and thus this result from RQME is to be trusted. Once again, the importance of carefully taking the accuracy of solutions of second order QME into consideration is seen.

\section{Calculations for the auxiliary cavity}\label{Sec:cavity_calculations}
We consider a cavity which is coupled to the DQD charge qubit via a Jaynes-Cummings type coupling, and also with its own thermal bosonic bath,
\begin{align}
&\hat{\mathcal{H}}_{c}= \omega_0 \hat{a}^\dagger\hat{a},~~\hat{\mathcal{H}}_{c-DQD}=g(\hat{\sigma}_+ \hat{a} +  \hat{a}^\dagger\hat{\sigma}_-), \nonumber \\
& \hat{\mathcal{H}}_B = \sum_{r=1}^\infty \Omega_r \hat{B}_r^\dagger \hat{B}_r, ~\hat{\mathcal{H}}_{c-B} = \sum_{r=1}^\infty \kappa_{r} (a^\dagger\hat{B}_r + \hat{B}_r^\dagger a) 
\end{align}
and $\hat{\sigma}_{\pm}=(\hat{\sigma}_x \pm i \hat{\sigma}_y)/2$.

Assuming the spectral density of the bosonic bath of the cavity is reasonably flat, we have equation of motion for the cavity field operator
\begin{align}
\label{cavity_eq1}
\frac{d\hat{a}}{dt}= -(i\omega_0+\kappa) \hat{a} - i \hat{\xi} -ig\hat{\sigma}_-,
\end{align}
where $\kappa$ gives the cavity decay rate, and $\hat{\xi}$ is the thermal `noise' operator coming from the bosonic environment having the properties
\begin{align}
&\langle\hat{\xi}(t)\rangle = 0,~~\langle\hat{\xi}^\dagger(t_1)\hat{\xi}(t_2)\rangle = \int_0^\infty \frac{d\omega}{2\pi} \kappa\mathfrak{n}_B(\omega) e^{i\omega(t_1-t_2)},~~\mathfrak{n}_B(\omega)=\frac{1}{e^{\beta\omega}-1}. 
\end{align}
The formally exact equation for motion for $\hat{\sigma}_-$ is given by
\begin{align}
&\hat{\sigma}_-(t) = \hat{\xi}_\sigma(t)+ig \int_0^t dt^\prime e^{-i(\omega_q+\Gamma)(t-t^\prime)} \hat{\sigma}_z(t^\prime) \hat{a}(t^\prime), \nonumber \\
&\hat{\xi}_\sigma(t)=\hat{\sigma}_-(t)|_{g=0} 
\end{align}
where we have introduced the parameter $\Gamma$, the inverse of which gives the time to reach steady state for the qubit in absence of the cavity.
Replacing this in Eq.~\eqref{cavity_eq1}, we have
\begin{align}
\frac{d\hat{a}}{dt}&= -(i\omega_0+\kappa) \hat{a} - i \hat{\xi} -ig\hat{\xi}_\sigma(t) +g^2 \int_0^t dt^\prime e^{-i(\omega_q+\Gamma)(t-t^\prime)} \hat{\sigma}_z(t^\prime) \hat{a}(t^\prime).
\end{align} 
Until this point, everything is exact. Now we make two approximations: (a) the cavity-DQD coupling is small, i.e, $g\ll \omega_0,\omega_q$, (b) the DQD-phonon coupling is also small, which translates into $\Gamma \ll \omega_q$. To keep track of the order of small terms, we introduce a small paramter $\epsilon$,
\begin{align}
g\rightarrow \epsilon g, ~~\Gamma\rightarrow \epsilon^2 \Gamma.
\end{align}
We will obtain our result up to $O(\epsilon^2)$. We will also assume that we are away from resonance and  $\omega_0<\omega_q$. We will also consider $t\gg \varepsilon^2\Gamma, \kappa$. With these approximations, we have up to $O(\epsilon^2)$,
\begin{align}
\frac{d\hat{a}}{dt}&= -(i\omega_0+\kappa) \hat{a} - i \hat{\xi}(t) -i\epsilon g\hat{\xi}_\sigma(t) + \epsilon^2 g^2  \hat{a}\langle\hat{\sigma}_z(t)\rangle\frac{\epsilon^2\Gamma-i(\omega_q-\omega_0)}{\epsilon^4\Gamma^2 + (\omega_q-\omega_0)^2}.
\end{align}  
Here we have used the fact that $\hat{\sigma}_z(t^\prime)=\hat{\sigma}_z(t)+O(\epsilon^2)$, $\hat{a}(t^\prime)=\hat{a}(t)e^{i\omega_0(t-t^\prime)}+O(\epsilon^2)$,  replaced $\hat{\sigma}_z(t)$ with its expectation value $\langle\hat{\sigma}_z(t)\rangle$, because  $\hat{\sigma}_z(t)$ is a conserved quantity to the leading order. For $t\gg \epsilon^2 \Gamma$, $\langle\hat{\sigma}_z(t)\rangle=\langle\hat{\sigma}_z\rangle$, where $\langle\hat{\sigma}_z\rangle$ is the steady state expectation value of $\hat{\sigma}_z$ in absence of cavity. So we have,
\begin{align}
\label{operator_eq}
&\frac{d\hat{a}}{dt}= -(i\omega_0^{eff}+\kappa^{eff}) \hat{a} - i \hat{\xi}(t) -i\epsilon g\hat{\xi}_\sigma(t) \nonumber \\
& \omega_0^{eff} = \omega_0 - \epsilon^2 g^2  \langle\hat{\sigma}_z\rangle_{ss}\frac{\omega_q-\omega_0}{\epsilon^4\Gamma^2 + (\omega_q-\omega_0)^2} \\
& \kappa^{eff}= \kappa + \epsilon^2 g^2  \langle\hat{\sigma}_z\rangle_{ss}\frac{\epsilon^2\Gamma}{\epsilon^4\Gamma^2 + (\omega_q-\omega_0)^2} \nonumber.
\end{align}
Away from resonance, i.e, for $\omega_q\neq \omega_0$, we have $\omega_{eff}\simeq \omega_0$ and $\kappa_{eff}\simeq \kappa$. Taking expectation value, we have
\begin{align}
\frac{d\langle\hat{a}(t)\rangle}{dt}&= -(i\omega_0+\kappa)\langle \hat{a}(t) \rangle  -i\epsilon g\langle \hat{\sigma}_-\rangle,
\end{align}
where we have used the fact that $\langle\hat{\xi}_\sigma(t)\rangle=\langle\hat{\sigma}_-(t)|_{g=0}\rangle = \langle \hat{\sigma}_-\rangle$ for $t\gg \epsilon^2 \Gamma$. Here, $\langle \hat{\sigma}_-\rangle$ is the steady state expectation value of $\hat{\sigma}_-$ in absence of cavity. This gives the steady state expectation value of the cavity-field operator as
\begin{align}
\langle\hat{a}\rangle= -\frac{\epsilon g}{\omega_0-i\kappa}\langle  \hat{\sigma}_-\rangle.
\end{align}
 In the steady state of the qubit, we have $\langle  \hat{\sigma}_-\rangle=\langle  \hat{\sigma}_x\rangle/2$. So, we get,
\begin{align}
\langle\hat{a}\rangle = -\frac{\epsilon g}{2(\omega_0-i\kappa)}\langle  \hat{\sigma}_x\rangle.
\end{align}

We can write down the formal solution of Eq.~\eqref{operator_eq},
\begin{align}
\label{formal_soln}
\hat{a}(t) =&     e^{-(i\omega_0+\kappa)t}\hat{a}(0)-i \int_0^{t} dt^\prime e^{-(i\omega_0+\kappa)(t-t^\prime)} \left(\hat{\xi}(t^\prime)+i\epsilon g \hat{\xi}_{\sigma}(t^\prime) \right).
\end{align}
The formal solution for $\hat{a}^\dagger(t)$ is given by the hermitian conjugate of the above equation.
Now, we note that 
\begin{align}
\langle\hat{\xi}^\dagger_{\sigma}(t_1)\hat{\xi}_{\sigma}(t_2)\rangle = \langle \sigma_+ \sigma_-\rangle e^{i\omega_q(t_1-t_2)} + O(\epsilon^2). 
\end{align}
Multiplying Eq.~\eqref{formal_soln} by its hermitian conjugate, taking expectation value and using above equation, we get the steady-state value of the cavity occupation,
\begin{align}
\langle \hat{a}^\dagger \hat{a} \rangle =  \epsilon^2 g^2 \frac{(1+\langle \hat{\sigma}_z \rangle)}{2(\kappa+i(\omega_0-\omega_q))} + \int_0^{\infty} \frac{d\omega}{\pi} \frac{\mathfrak{n}_B(\omega) \kappa}{(\omega-\omega_0)^2 + \kappa^2}. 
\end{align}
It can be checked using similar steps that $\langle \hat{a}^2\rangle =0$.
In the steady-state of the qubit, $\langle \hat{\sigma}_z \rangle=-1+O(\epsilon^2)$. This means, leading order contribution to cavity occupation from the DQD is zero. So, we have
\begin{align}
\langle \hat{a}^\dagger \hat{a} \rangle =  \int_0^{\infty} \frac{d\omega}{\pi} \frac{\mathfrak{n}_B(\omega) \kappa}{(\omega-\omega_0)^2 + \kappa^2}, 
\end{align}
which is essentially the occupation corresponding to the global thermal state of the cavity and the bosonic bath.  This guarantees that there is no current of photons between the cavity and the bosonic bath. This is consistent with the fact that the entire set-up is in thermal equilibrium. For small $\kappa$, $\langle \hat{a}^\dagger \hat{a} \rangle =\mathfrak{n}_B(\omega_0) $.

\end{document}